\documentclass[11pt]{article}

\usepackage{graphicx}
\usepackage{subfigure}
\usepackage{amsfonts,amssymb,amsmath}
\usepackage{vmargin}
\usepackage{cite}

\numberwithin{equation}{section}

\newcommand{\ka}{\ensuremath{\kappa}}

\newcommand{\N}{\ensuremath{{\cal N}}}

\newcommand{\half}{\ensuremath{\frac{1}{2}}}

\newcommand{\quarter}{\ensuremath{\frac{1}{4}}}

\newcommand{\be}{\begin{equation}}
\newcommand{\ee}{\end{equation}}

\newcommand{\ba}{\begin{eqnarray}}
\newcommand{\ea}{\end{eqnarray}}

\newcommand{\ns}{\normalsize}

\newcommand{\gsim}{\raise.3ex\hbox{$>$\kern-.75em\lower1ex\hbox{$\sim$}}}
\newcommand{\lsim}{\raise.3ex\hbox{$<$\kern-.75em\lower1ex\hbox{$\sim$}}}

\newcommand{\nn}{\nonumber}

\newcommand{\w}{\wedge}

\newcommand{\Vol}{\mathcal{V}}

\newcommand{\im}{\mathrm{Im\;}}
\newcommand{\re}{\mathrm{Re\;}}

\newcommand{\href}[1]{\underline{#1}}

\begin{document}

\begin{titlepage}
 
\title{
   \vskip 2cm
   {\Large\bf WEAKLY-coupled IIA Flux  Compactifications}
   \\[0.5cm]}
     \setcounter{footnote}{0}
\author{
{\ns\large 
  \setcounter{footnote}{3}
  Eran Palti$^1$\footnote{email: palti@thphys.ox.ac.uk}~, \;\;
  Gianmassimo Tasinato$^2$\footnote{email: gianmassimo.tasinato@uam.es}~, \;\;
  John Ward$^{3}$\footnote{email: jwa@uvic.ca}}
\\[0.5cm]
   $^1${\it\ns Rudulf Peierls centre for Theoretical Physics, University of Oxford}\\
   {\it\ns Keble Road, Oxford, UK. } \\[0.2em] 
   $^2${\it\ns Instituto de Fisica Teorica, UAM/CSIC
Facultad de Ciencias C-XVI,}\\
   {\it\ns C.U. Cantoblanco, E-28049-Madrid, Spain.} \\[0.2em] 
   $^3${\it\ns Department of Physics and Astronomy, University of Victoria, }\\
   {\it\ns Victoria, BC, V8P 1A1, Canada.}  
 }
 
\date{}
 
\maketitle
 
\begin{abstract}\noindent
We study compactifications of type IIA string theory on Calabi-Yau manifolds that are mirror to a subset of the type IIB LARGE-volume models. A combination of flux, $\alpha'$ corrections and non-perturbative effects stabilises the moduli in a non-supersymmetric AdS vacuum. This vacuum has the feature that the (ten-dimensional) string coupling is exponentially small which can naturally lead to a TeV supersymmetry breaking scale with an intermediate string scale. The AdS vacuum can be uplifted to a dS one through the introduction of D6 branes, and complex-structure modular inflation can be realised. 
\end{abstract}
 
\thispagestyle{empty}
 
\end{titlepage}

\tableofcontents

\section{Introduction}

Flux compactifications of type IIA and IIB string theories have been approached differently. In IIB it is known that, up to a conformal warp-factor, a Calabi-Yau (CY) solves the ten-dimensional supergravity equations of motion in the presence of fluxes \cite{Giddings:2001yu}. This is a nice property since we know much about CY manifolds. On the other hand, the K\"ahler moduli are not fixed by the fluxes and so must be fixed non-perturbatively \cite{Kachru:2003aw}. The presence of non-perturbative effects, although natural from a four-dimensional field-theory point of view, is difficult to implement into the full solution of the equations of motion\footnote{See for a possible approach \cite{Koerber:2007xk}.}. In that sense these vacua are not so well-understood. In IIA the fluxes have a  more drastic back-reaction and { \it generally} induce torsion on the manifold deforming it away from a CY \cite{Behrndt:2004km,Behrndt:2004mj,Lust:2004ig}. Therefore turning on the full fluxes implies that we should consider more general $SU(3)$-structure manifolds which, although we have some  examples \cite{Kachru:2002sk,Villadoro:2005cu,House:2005yc,Micu:2006ey,Ihl:2007ah,Grana:2006kf,Aldazabal:2007sn,Tomasiello:2007eq,Koerber:2008rx}, are extremely difficult to construct and generally not well understood. On the bright side though it is possible to fix the moduli completely perturbatively using fluxes and geometry only \cite{fluxrev}. This means that we have a better understanding of such solutions from a ten-dimensional point of view. So IIA flux compactifications have generally focused on these two key properties: perturbative moduli fixing and induced torsion \cite{fluxrev}. Of course the two theories are related by T-duality, but since this interchanges NS flux with geometry, identifying the duals to CY compactifications with NS flux leads to torsionful or non-geometric manifolds \cite{Kachru:2002sk,Gurrieri:2002wz,Shelton:2005cf,Aldazabal:2006up,Camara:2005dc,Benmachiche:2006df,Grana:2006hr,Micu:2007rd,Palti:2007pm,Wecht:2007wu} and we return to the problem  of how to construct them. 

In this paper we adopt a phenomenological approach to IIA flux compactifications. Rather than studying the more general solutions we look for vacua with attractive phenomenological features.     
In this area type IIB CY compactifications have been more successful and in particular the LARGE-volume models stand out \cite{Balasubramanian:2005zx,Conlon:2005ki,Conlon:2005jm}. Further there are a number of IIB inflation models (reviews can be found in \cite{McAllister:2007bg}) whilst in IIA we lack such constructions 
\cite{Hertzberg:2007ke,Hertzberg:2007wc}. The purpose of this paper is to partially bridge this gap in phenomenology by studying IIA flux compactifications that can recreate many
of the features of the IIB models. The natural way to approach this aim is by using mirror symmetry. However, as stated, this generally involves moving away from the well-understood CY compactifications. We can avoid this by turning on only particular NS fluxes. More explicitly, we can work at the `CY intersection' of the two theories where on either side only one component of NS flux is turned on. This is the electric component that can be thought of as having `no legs' along the three T-duality directions composing mirror symmetry. Further, it is possible for a CY to solve the IIA ten-dimensional equations of motion by restricting the fluxes and `smearing' the orientifold \cite{Acharya:2006ne}. Therefore we study IIA CY compactifications that have IIB CY mirrors with restricted fluxes.     

For these compactifications, at the cost of purely perturbative stabilisation, we find that the phenomenologically attractive IIB features can be consistently recreated. We find vacua where all the moduli are stabilised in an AdS or dS vacuum and supersymmetry is broken at the TeV scale where the hierarchy is generated dynamically without fine-tuning. These constructions also admit complex-structure modular inflation models. Perhaps the defining feature of our models is the fact that the string coupling $g_s$ is fixed at exponentially small values. These features are mirrors to the IIB LARGE-volume features and motivate the name WEAKLY-coupled.  Schematically, the models are constructed as follows. A combination of fluxes and $\alpha'$ corrections fix the K\"ahler moduli and dilaton in a perturbative regime. The complex-structure moduli are fixed using a combination of perturbative corrections away from the large complex-structure limit and the non-perturbative effects of gaugino condensation on D6-branes or E2-instantons. The resulting non-supersymmetric AdS vacuum can be uplifted to a de Sitter one by introducing D6-branes. 

The structure of the paper is as follows. In section \ref{cycomwithout} we derive the four-dimensional effective action for IIA string theory with fluxes on a CY orientifold at `tree-level' in $\alpha'$ but including corrections away from the large complex-structure limit.  
In section \ref{modstanoal} we study moduli stabilisation in these scenarios and find that the K\"ahler moduli cannot be stabilised at acceptable values. 
This is remedied in section \ref{secgencase} where we include $\alpha'$ corrections that, in combination with the fluxes, stabilise the K\"ahler moduli at acceptable values. 
We also show that the complex-structure moduli can be subsequently fixed in a non-supersymmetric WEAKLY-coupled AdS vacuum. 
In section \ref{upsect} we derive the D-terms induced by the introduction of D6-branes at angles to the O6-plane and show that they can be used to uplift the AdS vacuum to a de Sitter one.
We discuss some of the phenomenological features of our constructions in section \ref{Secdiscuss}.

\section{CY orientifold compactifications without $\alpha'$ corrections}
\label{cycomwithout}

In this section we derive the  $\N=1$ four-dimensional effective action resulting from a compactification of IIA string theory on a CY manifold in the presence of fluxes and 
orientifold six-planes ($O6$). We maintain the analysis at tree-level in  $\alpha'$, but include corrections away from the large complex-structure limit.
The contributions away from the large complex-structure limit are mirror to IIB $\alpha'$ corrections and play a crucial role in moduli stabilisation as studied in section \ref{secgencase}. 
In both the K\"ahler moduli and complex-structure moduli sectors we adopt the approach of starting with the $\N=2$ pure CY set-up and imposing the orientifold truncation constraints. This will 
be a useful approach when it comes to considering $\alpha'$ corrections in section \ref{secgencase}. 

\subsection{The K\"ahler moduli}

In a IIA compactification on a CY manifold the K\"ahler superfields $T^i$ arise from the expansion of the K\"ahler form $J$ and the NS two-form $B$ in the harmonic $h^{(1,1)}$ basis $\omega_i$
\be
B + iJ = \left( b^i + i\tau^i \right) \omega_i = T^i \omega_i \;.
\ee 
Their moduli space can be described by the use of a prepotential \cite{Candelas:1990pi}
\be
F = -\frac16 \frac{K_{ijk}T^iT^jT^k}{T^0} \;. \label{lvkpp}
\ee
We include a constant field $T^0$ (that we set to unity after differentiation of the prepotential) playing the role of the mirror to the complex-structure homogeneous parameter, and 
 we  introduce the capital index $I=\{0,i\}$. The corresponding K\"ahler potential is obtained by the formula
\be
K^T \,= \, - \mathrm{ln\;} i\left[ \bar{T}^I F_I - T^I \bar{F}_I  \right] \,\equiv\, - \mathrm{ln\;} 8\Vol \;, 
\ee
where $F_I \equiv \partial_{T_I} F$, and  we introduce the quantities
\be
\Vol \equiv \frac16 K_{ijk} \tau^i \tau^j \tau^k \equiv \frac16 \kappa \hskip0.5cm \;, 
\hskip0.5cm \kappa_i \equiv K_{ijk}\tau^j\tau^k  
 \hskip0.5cm \;,\; 
\hskip0.5cm 
 \kappa_{ij} \equiv K_{ijk} \tau^k \;.
\ee

\smallskip

The orientifold truncation of this sector acts
simply  by  reducing  the index range of the fields so that we keep 
only the expansion  in the  odd
 two forms \cite{Grimm:2004ua}. This does not modify the structure
of the quantities we have introduced,  
 and so we maintain the same index labels for the truncated spectrum. For later use we display the K\"ahler derivatives
\be
K^T_{i\bar{j}} = -\frac{3}{2\kappa} \left( \kappa_{ij} -  \frac{3\kappa_i \kappa_j}{2\kappa} \right) \equiv  -\frac{3}{2\kappa} g_{ij}  \hskip0.5cm, \hskip0.5cm (K^T)^{i\bar{j}} = -\frac{2\kappa}{3} \left( \kappa^{ij} - \frac{3\tau^i\tau^j}{\kappa} \right) \equiv -\frac{2\kappa}{3} g^{ij}\;.
\ee

\smallskip

Another quantity that plays a role in our calculations is the `gauge coupling' matrix $N$ which is
given by the well known  $\N=2$ formula  (see \cite{Andrianopoli:1996cm,Grimm:2004ua} for more details) 
\be
N_{IJ} = \bar{F}_{IJ} + \frac{2i \left( \im{F} \right)_{IK}T^{K}  \left( \im{F} \right)_{JL}T^{L} }{\left( \im{F} \right)_{MN}T^{M}T^{N} } \label{gkpt} \;,
\ee
where ${F}_{IJ}\equiv \partial_{T_I} \partial_{T_J} F$. 
Its explicit form for the prepotential (\ref{lvkpp}) is given in Appendix \ref{AppendixA}.

\subsection{The complex-structure moduli}
\label{csnoal}

In IIA CY compactifications the complex-structure moduli arise from an expansion of the unique holomorphic three-form $\Omega$ in the harmonic three-form basis. 
In the presence of an $O6$, following \cite{Grimm:2004ua}, we split this expansion into an even and odd basis
\be
\left ( \alpha_{\hat{k}}, \,\beta^{\lambda} \right) \in H^3_+ \;\hskip0.4cm, \hskip0.4cm
\; \left ( \alpha_{\lambda}, \,\beta^{\hat{k}} \right) \in H^3_- \;\;.
\ee
Here the index range is such that summing over the indices  $\hat{k}$ and $\lambda$ gives the total number of real harmonic three-forms $2(h^{(2,1)}+1)\,$. 
The index $\hat{k}$ is defined in the range $\{0,k\}$. Choosing the $\alpha_0$ form to be even corresponds to the mirror  of  $O3/O7$ (rather than $O5/O9$) compactifications on the IIB side.  

The chiral superfields of the truncated $\N=1$ theory arise as in \cite{Grimm:2004ua}. We define the compensator field  $C$ as
 \be
C \,\equiv\, e^{-i\theta}e^{-D}e^{\half K^{cs}} \;. \label{dilrel}
\ee 
Here $\theta$ is a constant angle whose value is
set by the orientifold. $D$ is the four-dimensional dilaton which is given in terms of the ten-dimensional dilaton $\hat{\phi}$ by the relation
\be
e^{2D} \,= \, \frac{e^{2\hat{\phi}}}{\Vol} \;.
\ee 
The ($\N=2$) complex-structure K\"ahler potential, $K^{cs}$,  is given by
\be
K^{cs} \,= \,- \mathrm{ln}\; i\int_{CY}{\Omega \w \bar{\Omega} } \;.
\ee
The expansion of the three-form reads
\be
C\Omega \,=\, \re{(CZ^{\hat{k}})} \alpha_{\hat{k}} + i \im{(CZ^\lambda)} \alpha_\lambda - \re{(CF_{\lambda})} \beta^{\lambda} - i\im{(CF_{\hat{k}})} \beta^{\hat{k}} \;. \nn \\
\ee
This arises from the $\N=2$ expansion of the $\Omega$ 
form, after imposing the
 orientifold constraints \cite{Grimm:2004ua}
\be
\im{(CZ_{\hat{k}})} = \re{(CF_{\hat{k}})} = \re{(CZ_\lambda)} = \im{(CF_\lambda)} = 0 \;. \label{orcon}
\ee 
The RR three-form transforms evenly, and so can be  expanded as
\be
C_3 = \xi^{\hat{k}} \alpha_{\hat{k}} - \tilde{\xi}_{\alpha} \beta^{\alpha} \;.
\ee

\noindent
The complex chiral superfields of the resulting $\N=1$ theory are given by the expressions
\ba
S \equiv s + i\sigma  &=&  2 \re{(CZ^0)} - i \xi^0 \;, \\
U_{\lambda} \equiv u_{\lambda} + i \nu_{\lambda} &=& - 2 \re{(CF_{\lambda})} + i\xi_{\lambda}  \;, \\
N^k &=&2 \re{(CZ^k)} - i \xi^k  \;,
\ea
and the corresponding chiral K\"ahler potential reads  \cite{Grimm:2004ua}
\be
K^Q = 4D = - 2 \mathrm{ln\;} \left[ 2 \left( \re{(CF_{\lambda})}\im{(CZ^{\lambda})}  - \re{(CZ^{\hat{k}})} \im{(CF_{\hat{k}})}  \right) \right] \;. \label{genKQ}
\ee

\bigskip

We can write more explicit expressions for these quantities, rendering more manifest mirror symmetry with the IIB $O3/O7$ compactifications.  
In particular we would like to derive the mirrors of the IIB $\alpha'$ corrections and these should correspond to corrections away from the large complex-structure limit. 
To derive these we again consider the explicit form of the $\N=2$ prepotential away from the large complex-structure limit  as derived in \cite{Candelas:1990rm}
\be
F = \frac{1}{6}d_{abc}\frac{Z^a Z^b Z^c}{Z^0} + d^{(1)}_{ab}Z^{a}Z^{b} - \half d^{(2)}_{a}Z^{a}Z^0 - i (Z^0)^2 \xi + {\cal O} \left( e^{iZ} \right)\;. \label{mriibp}
\ee
Here $d_{abc}$, $d^{(1)}_{ab}$, $d^{(2)}_{a}$ are rational coefficients and $\xi$ is a real number. The co-ordinates $Z^A = \{Z^0,Z^a\}$ are homogenous co-ordinates of  the complex-structure moduli  $z^a$ of the CY. From here on we neglect the exponentially suppressed corrections that appear in the last term of (\ref{mriibp}).

We now introduce the O6 planes. The index structure splits as $A=\{0,k,\lambda\}$ and in order to match the usual IIB expressions we choose the symplectic basis $N^k=0$  (keeping only the component $N^0=S$). Imposing the orientifold constraints (\ref{orcon}) we find  $d^{(1)}_{\lambda\rho}=0$, and the resulting K\"ahler potential reads 
\ba
K^{Q} &=& - 2 \mathrm{ln\;} \left\{ 2 \left[ \re{(CF_{\lambda})}\im{(CZ^{\lambda})}  - \re{(CZ^0)} \im{(CF_0)}  \right] \right\} \; \nn \\
&=&  - \mathrm{ln\;} \left[ S + \bar{S} \right] - 2\,\mathrm{ln\;} f(q) -  \mathrm{ln\;}2\;.  \label{KQ}
\ea
The function $f(q)$ is defined by 
\ba
f(q) &\equiv& \frac16 d_{\lambda\rho\sigma} q^\lambda q^\rho q^\sigma + s^{3/2} \frac{\xi}{2} \equiv \Vol' + \frac{\xi' }{2}\;, \\
q^{\lambda} &\equiv& - 2 s^{-\half} \im{(CZ^\lambda)}  \;.
\ea 
In terms of $f(q)$ we have the superfields expression
\be
u_\lambda = \partial_{q^\lambda}f  \;.
\ee

In general the K\"ahler potential satisfies a no-scale like condition \cite{Grimm:2004ua}
\be
K^{Q}_{N} \left(K^{Q}\right)^{N\bar{M}} K^Q_{\bar{M}} = 4 \;,
\ee
where the index $N$ runs over all the superfields $N=\{S,U_{\lambda}\}$. 
In the absence of the $\xi$ parameter the dilaton and complex-structure contributions decouple and then we recover the exact no-scale condition 
\be
K^{Q}_{u_\lambda} \left(K^{Q}\right)^{u_\lambda\bar{u}_\sigma} K^Q_{\bar{u}_\sigma} = 3 \;. \label{csnosc}
\ee 

This set-up is exactly the mirror to IIB on a CY with $O3/O7$ planes. In our case the $q^{\lambda}$ play the role of the IIB (string frame) two-cycle volumes, and $u_{\lambda}$ correspond to the four-cycle volumes. $\Vol'$ is the mirror to the IIB CY volume. 
The $\xi$ term accounts for $\alpha'^3$ correction terms in the mirror IIB side  (as shown in the $\N=2$ case in \cite{Candelas:1990rm}).
Notice that the IIA complex-structure prepotential includes perturbative terms in $d^{(1)}$ and $d^{(2)}$ corresponding to IIB lower order $\alpha'$ corrections. These cancel out of the K\"ahler potential, and also do not appear in the superpotential: consequently they do not affect the theory. The exponentially suppressed terms in the IIA complex-structure prepotential correspond to IIB world-sheet instantons. 

\subsection{The fluxes and  superpotential}

Having described the structure of the K\"ahler potential, we turn to the superpotential induced by  the presence of fluxes. 
We choose to   turn on the following  fluxes
\be
H = -h_{0} \beta^{0} \;, \;\; F_0 = - f_0 \;, \;\; F_2 = - \tilde{f}^i \omega_i \;, \;\; F_4 = - f_i \tilde{\omega}^i \;, \;\; F_6 = -  \tilde{f}_{0} \epsilon \;. \label{fluxes}
\ee
We switch on {\it only one} component of $H$-flux. 
This is the electric component that threads the cycle orthogonal to the one wrapped by the $O6$ that is the mirror to the IIB $O3$ . In going to the mirror this can be thought of as the component that has `no-legs' along the three T-duality directions. It   gets mapped to $H$-flux on the IIB mirror. Other $H$-flux components would get mapped into non-geometric fluxes in a $O3/O7$ IIB set-up \cite{Grana:2006hr,Micu:2007rd}. 
The fluxes induce the superpotential \cite{Grimm:2004ua}
\be
W =  \frac{f_0}6 K_{ijk}T^i T^j T^k + \half K_{ijk} \tilde{f}^i T^j T^k - f_i T^i + \tilde{f}_0 - i h_0 S \;.
\ee
The absence of complex-structure superfields $U_{\lambda}$ in the superpotential, combined with (\ref{csnosc}), 
implies  that in the absence of the $\xi$ term the compactification is no-scale.  

\bigskip

The values of the fluxes are constrained by the tadpole equations \cite{DeWolfe:2005uu}
\ba
-f_0 h_{0}  &=& Q_{0} = 2N^{D6}_0 - 4 N^{O6}_0  \;\;, \\
0 &=& Q^{\lambda} = 2\left(N^{D6}\right)^{\lambda} - 4 \left(N^{O6}\right)^{\lambda}  \;. \label{tadpoles} 
\ea
Here $Q_0$ corresponds to the local charge induced by $D6$ and $O6$ planes\footnote{Recall that the charge of an $O6$ is $-4$ with respect to a $D6$ and that we must include the orientifold images of the $D6$ branes.} wrapping $\alpha_0$. 
The charges $Q^{\lambda}$ correspond to the cycles $\beta^{\lambda}$. We see that for those cycles the local sources must cancel by themselves.
The local charge $Q_0$ will play an important role in the analysis, since it  will constrain the size of the vevs of the moduli.
It is therefore worth pointing out  that it can take values up to $10^2-10^3$ \cite{Denef:2004dm}. 
Note that the constraints are independent of most of the fluxes which follows from the fact that the tadpole constraints are generally products of RR fluxes and NS or metric fluxes. 
Since we only have $h_{0}$ turned on, this choice substantially reduces the number of constraints.
The absence of the fluxes from the tadpoles does not imply a formally infinite number of solutions, since some choices of fluxes are related by axion shifts, making them physically equivalent.
 
\bigskip

In this paper we consider CY compactifications with fluxes turned on. This corresponds to neglecting the back-reaction of the fluxes on the geometry which is sometimes referred to as the `CY with fluxes' limit. 
In the IIB case the back-reaction induces a warp-factor  which only deforms the CY conformally \cite{Giddings:2001yu} and, in the large volume approximation,  does not change its key properties.  In IIA the back-reaction can be more drastic, deforming the CY to a half-flat manifold \cite{Behrndt:2004km,Behrndt:2004mj,Lust:2004ig}. However it has been 
shown in \cite{Acharya:2006ne}, that turning off the fluxes $F_2$ and $F_6$, and additionally 'smearing' the orientifold,  an unwarped CY compactification still solves the 
complete ten-dimensional equations of motion. As we will see,  setting the fluxes $\tilde{f}_0=\tilde{f}^i=0$ does not qualitatively affect the  properties of our solutions. Therefore, although we will consider general fluxes for the sake of completeness, we can keep in mind that the more accurate scenario of \cite{Acharya:2006ne}  can be reached as a suitable limit of our constructions without affecting the results.
      
\section{Moduli stabilisation  without  $\alpha'$ corrections}
\label{modstanoal}

Having established our set-up, we go on to study moduli stabilisation within this scenario. 
The $\N=1$ effective theory is specified by the K\"ahler potential and the superpotential
\ba
K &=& K^T(T) + K^Q(S,U) = - \mathrm{ln\;} 8\Vol - \mathrm{ln\;} \left[  S + \bar{S}  \right]  - 2\mathrm{ln\;} \left[\Vol' + \frac{\xi' }{2}\right]  -  \mathrm{ln\;} 2 \;, \nn \\ 
W &=&  W^T(T) + W^Q(S) =  \frac{f_0}6 K_{ijk}T^i T^j T^k + \half K_{ijk} \tilde{f}^i T^j T^k - f_i T^i + \tilde{f}_0 - i h_0 S \;. \label{kqms}
\ea 
where the definitions of the various quantities are provided in section \ref{cycomwithout}. 
 
Our aim is  to recreate, in this type IIA context, the moduli stabilisation constructions that have been developed on the type IIB side. 
There, at tree-level, the K\"ahler moduli sector is characterised by a no-scale structure while the complex-structure moduli are fixed by fluxes  \cite{Giddings:2001yu}. 
The additional inclusion of non-perturbative effects, and $\alpha'$ corrections, allows to fix also the K\"ahler moduli \cite{Kachru:2003aw,Balasubramanian:2005zx}. 
The situation in our IIA set up can be considered as a mirror to the IIB scenario just described, with all but one of the NS flux parameters turned off. As in IIB, we study moduli stabilisation using 
a two-stage procedure where we first consider the no-scale vacuum and solve for the IIA K\"ahler moduli and dilaton F-terms. This corresponds to neglecting the $\xi$ term in (\ref{kqms}) and not considering non-perturbative effects in which case the complex-structure moduli are characterised by a no-scale structure and decouple from the theory.  
The second step is to fix the complex-structure moduli but, as we proceed to show, already in the first step we encounter problems.

\subsection{Solving the K\"ahler moduli F-terms}
\label{SecFtermsim}

We proceed to  solve the dilaton and K\"ahler moduli F-terms. An analysis similar to ours was performed in \cite{DeWolfe:2005uu}. 
There it was shown that if all the H-flux components are turned on,  then all the complex-structure and 
K\"ahler moduli can be fixed perturbatively. In our case we are interested in turning on only one H-flux component and instead stabilising the 
complex-structure moduli non-perturbatively. However,  the choice of H-flux turns out to also influence the stabilisation of the K\"ahler moduli sector.
Indeed we will see that the K\"ahler moduli are either fixed at unphysical values (where the volume of the internal manifold vanishes), or are left as flat directions. 
This may appear counter intuitive, especially since in the analysis of \cite{DeWolfe:2005uu} the F-term
conditions for the K\"ahler and complex-structure moduli decouple and are solved independently. However, the particular decoupling found in \cite{DeWolfe:2005uu} relied on the fact that 
all the H-flux is switched on. 
     
\bigskip
          
Using (\ref{kqms}) the F-term for the dilaton reads 
\be
D_SW = 0 = W_S + K_S W = -i h_{0} - \frac{W}{S+\bar{S}} \;,
\ee
which fixes the dilaton and its axion as
\be
-i h_{0} \bar{S} = W^T \;. \label{SF}
\ee 
When this is satisfied we have the nice property that  
\be
W = 2i \im{(W^T)} \;, \label{ussim}
\ee
which means that the F-terms for the K\"ahler moduli are independent of the dilaton and can be written as
\be
W^T_{T^i} + 2i K_{T^i} \im{(W^T)} = 0 \;. \label{TFT}
\ee
This crucially differs from a similar expression in \cite{DeWolfe:2005uu} by a factor of $-2$ in the second term which arises because of the different choice of H-flux.
The imaginary part of (\ref{TFT}) fixes the axions as
\be
b^i = -\frac{\tilde{f}^i}{f_0} \;.  \label{bvev}
\ee
The real part reads
\ba
&\;& \half f_0 \kappa_{ijk} ( \tau^j \tau^k - b^j b^k ) - \kappa_{ijk} \tilde{f}^j b^k + f_i \nn \\
&-& \frac{3\kappa_{i}}{\kappa} \left[ \frac16 f_0 \kappa_{lmn} \left( \tau^l \tau^m \tau^n - 3 \tau^l b^m b^n \right) - \kappa_{lmn}\tilde{f}^l \tau^m b^n + f_l \tau^l \right] = 0 \label{TFT2} \;.
\ea
Substituting (\ref{bvev}) into (\ref{TFT2}) and multiplying by $4f_0 \kappa$ gives
\ba
&\;& \kappa \left[ 2 f_0^2 \kappa_{i}+ 2 \kappa_{ijk}\tilde{f}_j\tilde{f}_k + 4f_0 f_i \right] \nn \\
&-& 2 \left[  \kappa f_0^2 \kappa_{i} + \kappa_{i} \left( 3 \kappa_{lmn} \tilde{f}_l \tilde{f}_m \tau^n + 6 f_0 f_l \tau^l  \right) \right] = 0 \;. \label{TFT3}
\ea
Contracting with $\tau^i$  we get
\be
\kappa \tau^i \left[  \kappa_{ijk}  \tilde{f}_j \tilde{f}_k + 2 f_0 f_i  \right]   = 0 \;. \label{TFT4}
\ee
Then substituting (\ref{TFT4}) into (\ref{TFT3}) we arrive at 
\be
\kappa \left[ \kappa_{ijk} \tilde{f}_j \tilde{f}_k + 2 f_0 f_i  \right] = 0 \;. \label{TFT5}
\ee 
For general fluxes, this condition fixes the moduli at vanishing volume. This is unacceptable and so we are forced  to pick the fluxes so that the term in the brackets vanishes. But
this in turn is just a constraint on the flux parameters, and so none of the K\"ahler moduli get fixed\footnote{We emphasise that the moduli are only flat directions in the case where we neglect the effects of the $\xi$ parameter. Once it is included,  the no-scale property of the complex structure sector will be broken,  and, as shown in section \ref{scakpot}, the K\"ahler moduli become runaway rather than flat directions.}. Therefore
with this choice all the moduli other than the dilaton remain as flat directions \footnote{Note that this is only possible because supersymmetry is broken, 
since the axion partners of the K\"ahler moduli \emph{do} get fixed.}. To see how the dilaton is fixed we substitute (\ref{TFT5}) into (\ref{SF}) and recover
\ba
s &=& \frac16 \frac{f_0}{h_0} \kappa \nn \;, \\
\sigma &=& -\frac{1}{h_0f_0} \left(  \frac13 \tilde{f}_i f_i + \tilde{f}_0 f_0 \right) \;. \label{snoal}
\ea
This analysis exactly agrees with the special case in \cite{Camara:2005dc} where a no-scale model on a torodial orbifold was 
studied.

\bigskip
 
The IIB mirror to the analysis of the dilaton and K\"ahler moduli F-terms is the Imaginary Self Dual (ISD) condition on the flux \cite{Giddings:2001yu}. In 
Appendix \ref{AppendixA}  we show indeed that this condition, in the mirror IIB side, exactly leads to the same solution as the one we find in IIA. 

\subsection{The scalar potential}
\label{scakpot}

It is interesting to examine these results at the level of the scalar potential.  The scalar potential reads \cite{Grimm:2005fa}
\be
V = \frac{9e^{2\hat{\phi}}}{\ka^2} \int{H_3 \w \star H_3} - \frac{18\,e^{4\hat{\phi}}}{\kappa^2} \left( \tilde{e}_I - N_{IJ} \tilde{m}^{J} \right) \left( \im{N}^{-1}\right)^{JK}  \left( \tilde{e}_K - \bar{N}_{KL} \tilde{m}^{L} \right) + V_{O6}\;, \label{scapot}
\ee
where  the gauge coupling matrix $N$ was introduced in equation (\ref{gkpt}). 
In terms of the fluxes (\ref{fluxes}) we have 
\ba
\tilde{e}_I &=& \left( \tilde{f}_0  - \xi^0 h_0 , -f_i \right) \hskip0.5cm ,\hskip0.5cm \tilde{m}^I = \left( -f_0, \tilde{f}_i \right) \;.
\ea
We have added to the expression in \cite{Grimm:2005fa} the local orientifold contribution $V_{O6}$.  This reads (see  \cite{Villadoro:2005cu,DeWolfe:2005uu} and section \ref{upsect} for a derivation)
\be
V_{O6} =  f_0 e^{4D}  \im{W^Q}  \;.
\ee
We want to eliminate the ten and four-dimensional dilatons for the dilaton superfield $S$. They are connected through (\ref{dilrel}), which provides
\be
s = 2\re{\left( CZ^0 \right)} =2 e^{\half K_{cs} }e^{-D} \re{\left(e^{-i\theta} Z^0 \right)} \;.
\ee
Then the resulting scalar potential reads 
\be
V = R \left[  \frac{3h_0^2r}{2\ka s^2} - \frac{1}{2s^4} \left( \tilde{e}_I - N_{IJ} \tilde{m}^{J} \right) \left( \im{N}^{-1}\right)^{JK}  \left( \tilde{e}_K - \bar{N}_{KL} \tilde{m}^{L} \right) -  \frac{f_0 h_0 }{s^3} \right] \;,
\ee
where we have defined the quantities
\ba
R &\equiv& 16e^{2K_{cs}} \re{\left(e^{-i\theta} Z^0 \right)}^4   \;, \\
r &\equiv& \frac{e^{-K_{cs}}}{4\re{\left(e^{-i\theta} Z^0 \right)}^2 } \int{\beta^0 \w \star \beta^0} =  - \frac{e^{-K_{cs}}}{4\re{\left(e^{-i\theta} Z^0 \right)}^2 } \left( \im{(M)}^{-1}\right)^{00} \nn \\
 &=&  \frac{ \left ( \Vol' +  \frac{\xi'}{2} \right) }{ \left( \Vol' - \xi' \right) \left[ 1 - 2\frac{\Vol' - \xi'}{4\Vol' - \xi'} \right]  }  \;\,\,.
\ea
The matrix $M$ assumes the same form as $N$ in (\ref{gkpt}), with the complex-structure moduli replacing the K\"ahler moduli (in other
words  we send $X^i \rightarrow Z^{\lambda}$),
and additionally imposing the orientifold constraints (\ref{orcon}) \cite{Grimm:2004ua}.
The case $r=2$ corresponds exactly to  the large complex-structure or no-scale limit.

\subsubsection{Reproducing the solution}

We now proceed to recover the solution obtained  from the F-term  analysis in section \ref{SecFtermsim}. To do so
we set $r=2$ since this is the no-scale case studied. Away from this limit there are corrections that will be suppressed by powers of $\Vol'$,  
which we discuss in section \ref{secgencase}.  
To simplify the analysis we set the fluxes $\tilde{f}_i=\tilde{f}_0=0$ and, matching the F-terms solutions, take $b^i=\sigma=0$. 
This gives\footnote{This can be checked by direct dimensional reduction, using the supergravity formula for the scalar potential and the relation $e^K = \frac{3R}{4\kappa s^4}$.}
\ba
V &=& R \left[  \frac{3h_0^2}{\ka s^2} + \frac{\kappa f_0^2}{12 s^4}  - \frac{f_0 h_0 }{s^3} - \frac{1}{2s^4} g^{ij}f_if_j    \right] \;.
\ea
The F-term equations are equivalent to the minimum equations for this potential, and we can see this by considering  the combination
\be
4\tau^i \partial_{\tau^i}V - s\partial_sV = \frac{R}{3s^4\kappa}\left[ -90 h_0^2 s^2 + 4 \kappa^2 f_0^2 - 9 f_0 h_0 \kappa s \right] \;,
\ee
which is solved by 
\be
s = \frac{f_0}{6h_0} \kappa \;. \label{ssolnoa}
\ee
This is exactly the F-terms solution (\ref{snoal}). Now using this solution we can write 
\be
\tau^i  \partial_{\tau^i}V = \frac{R}{2s^4} g^{ij} f_i f_j =  \frac{R}{2s^4} |f|^2 \;, \label{condnofix}
\ee
where we used $\tau^i \partial_i g^{jk} = -g^{jk}$. This is the analogous condition to (\ref{TFT4}) which forces us to choose $f_i=0$ 
otherwise the K\"ahler moduli are fixed in a non-physical regime. If we take $f_i=0$ then the K\"ahler moduli become flat directions 
since then (\ref{ssolnoa}) implies $\partial_{\tau^i}V=0$. 
Note that for the case $r\neq 2$ there are no solutions to $\partial_{\tau^i}V=0$ which shows that they destabilise.

\subsubsection{Relation with the no-go theorem of  \cite{Hertzberg:2007wc}}

There is a further interesting property of the $r=2$ no-scale solution in relation to the no-go theorem of  \cite{Hertzberg:2007wc}. There the scalar potential was decomposed as a sum of positive definite terms called $V_p$, with $p$ denoting the degree of the flux that gives rise to that term, and positive and negative contributions from D6 branes and O6 planes respectively \cite{Hertzberg:2007wc}
\ba
V &=& V_3 + V_0 + V_2 + V_4 + V_6 + V_{D6} - V_{O6} 
\nonumber \\
&=& \frac{A_3}{\tau^3s^2}  + \frac{A_0\tau^3}{s^4} + \frac{A_2\tau}{s^4} + \frac{A_4}{\tau s^4} + \frac{A_6}{\tau^3s^4} +  \frac{A_{D6}}{s^3}-  \frac{A_{O6}}{s^3} \;. \label{kachsca}
\ea
In \cite{Hertzberg:2007wc}
it was shown that neutrally-stable Minkowski vacua require the condition  $V_2=V_4=V_6=0$.   
Since we have set $F_2=F_4=F_6=0$ it is immediate to see that the $r=2$ case is such a  Minkowski vacuum where the (trivial) tadpoles are satisfied. 
We can also analyse the more general case  where all the RR fluxes are kept on. To do this, we use the identities
\ba
\left( \tilde{e}_0 - \re{N}_{0J} \tilde{m}^{J} \right) &=& \tilde{f}_0+ h_0 \sigma-\frac{1}{3} K_{ijk}  b^i b^j b^k f_0 - \frac12 K_{ijk} b^j b^k \tilde{f}^i = 0 \;, \nn \\
\left( \tilde{e}_i - \re{N}_{iJ} \tilde{m}^{J} \right)  &=& -f_i +\frac{1}{2} K_{ijk}b^j b^k f_0 + K_{ijk} b^j \tilde{f}^k = 0\;,
\ea
that are straightforward to prove using the solutions for $\sigma$ and $b^i$, formulae  (\ref{Nnoal}) for the  $N$ matrix,
and imposing the constraint on the fluxes that comes from setting to zero the term in brackets in formula (\ref{TFT5}).  
This gives  
\ba
V  &=& -\frac{R}{2s^4} \left( \tilde{e}_I - N_{IJ} \tilde{m}^{J} \right) \left( \im{N}^{-1}\right)^{IK}  \left( \tilde{e}_K - \bar{N}_{KL} \tilde{m}^{L} \right) = \frac{Rk}{12s^4}f_0^2 \,=\,V_0\;, \\
V_2 &=& V_4 \,=\, V_6 \,=\,0 \;,
\ea
as required. So we see that the no-scale case relates nicely to the no-go theorem of \cite{Hertzberg:2007wc}: there it was argued that the condition $V_2=V_4=V_6=0$ is difficult to satisfy due to the tadpole constraints. In this case the tadpoles do not constrain the fluxes but the F-terms do.

\section{Moduli stabilisation with $\alpha'$ corrections}
\label{secgencase}

We have seen that the K\"ahler moduli either get stabilised at unphysical values or are left as flat directions.
A possible  way to avoid this is to move away from a CY compactification by introducing metric or non-geometric fluxes.
However, in moving away from CY compactifications, we lose explicit control over our  constructions. For example, mirror symmetry predicts that there should be half-flat manifolds
for which, away from the large complex-structure limit, the corrections to the complex-structure K\"ahler potential have the form as in (\ref{KQ}). 
However these corrections have not been calculated explicitly.  
    
To maintain explicitness we consider only CY compactifications. We focus on the effects of $\alpha'$ corrections. Naively, it seems unlikely that such $\alpha'$ corrections can stabilise the K\"ahler moduli at acceptable values whilst still maintaining the validity of the $\alpha'$ expansion: recall that the tree-level potential (henceforth we refer to the leading terms in $\alpha'$ as tree-level) pushes the K\"ahler moduli towards an unphysical regime. In order to move them substantially away from this region we must ensure that  $\alpha'$ corrections compete with the tree-level result. This seems to violate the $\alpha'$ expansion. 
Fortunately, in the presence of fluxes, we will see that it is possible to have a well-controlled $\alpha'$ expansion and still have tree-level terms in one type of flux competing with higher order terms in a different type of flux. We reserve a more detailed discussion of this to section \ref{algsexp}. 

This section begins with a derivation of  $\alpha'$ corrections in our case. We then go on to study their effects on moduli stabilisation and show that they allow to fix the dilaton and K\"ahler moduli
at acceptable values. In section \ref{scalpotgenc} we show how this is realised in the scalar potential. In section \ref{algsexp}  we discuss the validity of the $\alpha'$ expansion and the effects of other corrections to the potential. In section \ref{csmoduli} we complete the moduli stabilisation framework by including the complex-structure moduli which are stabilised by a combination of the $\xi$ term and non-perturbative effects. 

\subsection{The  $\alpha'$ corrections}

Deriving $\alpha'$ corrections is a very difficult task. Few of these corrections can be calculated explicitly. In \cite{Becker:2002nn} a correction associated with the internal gradient of the dilaton, induced by the $\alpha'^3 R^4$ correction to the type II supergravity action, was obtained. This corresponds to a correction to the volume factor that multiplies the four-dimensional Ricci scalar, following integration over the internal space
\be
\Vol R_{4} \rightarrow \left( \Vol + \frac{\epsilon}{2} \right) R_{4} \;,
\label{pralpr}
\ee
where $\epsilon = -\frac{\chi \zeta(3)}{2}$ with $\chi$ being the CY Euler number. 
Starting from this explicitly calculated correction, we can use supersymmetry arguments to infer the form of other corrections, since they all must fit into a supergravity formalism. 
It was shown in \cite{Becker:2002nn} that (\ref{pralpr}) can be accommodated in supergravity by appropriately modifying the K\"ahler potential. Furthermore, it was shown that it is exactly the correction that arises from the (orientifold truncated) modification of the prepotential predicted by mirror symmetry \cite{Candelas:1990rm} (in our notation it corresponds to the mirror 
of the $\xi$ term in (\ref{KQ})). Once we know the correction to the prepotential/K\"ahler potential, we can deduce corrections other than (\ref{pralpr}) by using the supergravity formula for the scalar potential. This is indeed the approach taken in \cite{Balasubramanian:2005zx} where the key corrections in that scenario comes from  $\alpha'^3 R^3 H^2$ type terms.
These were not derived in \cite{Becker:2002nn} but rather inferred using supersymmetry arguments. 

In this paper we adopt the same approach. We take the prepotential to be of the form\footnote{The prepotential (\ref{corpot})  receives 
further exponentially suppressed worldsheet instanton corrections,
 which we neglect.} 
\be
F = -\frac16 \frac{K_{ijk}T^iT^jT^k}{T^0} + K^{(1)}_{ij}T^iT^j + K^{(2)}_i T^i T^0 - i \epsilon (T^0)^2 \;. \label{corpot}
\ee
The first term is the tree-level result, and the $\epsilon$ term is the term that is required to account for the volume correction. We also include two new corrections $K^{(1)}$ and $K^{(2)}$. These are not required to account for the volume factor, but are consistent with it since they drop out of the K\"ahler potential. 
Instead they are required by mirror symmetry, as we should match the prepotential (\ref{mriibp}), and
 correspond to lower order $\alpha'$ corrections. Notice
 that such a correction does not occur for the pure CY case, and this is reflected in the fact that they drop out of the K\"ahler potential; 
however, in the presence of fluxes they do modify the theory. 
 
In the IIB case, supersymmetry arguments provide the form of corrections to the K\"ahler potential. In the IIA case, they also suggest the form of the  corrections to the superpotential. Indeed,
while the IIB superpotential is unaffected by $\alpha'$ corrections, being protected by shift symmetries of the RR axions that superpartner the K\"ahler moduli \cite{Dine:1986vd}, in IIA there are no such symmetries and the superpotential normally receives $\alpha'$ corrections\footnote{$\alpha'$ corrections are not interpreted in the effective four-dimensional theory as quantum corrections thereby satisfying more general non-renormalisation theorems.}.

The orientifold acts on the prepotential as a truncation of the index range, and so the effective $\N=1$ theory we are going to study is given by
\ba
K^T &=&  - \mathrm{ln\;} 8\left( \Vol + \half \epsilon \right) \;. \nn \\
W^T &=&  f_0 F_0 - \tilde{f}^i F_{i} - f_i T^i + \tilde{f}_0  \nonumber \\&=&
  \frac{f_0}6 K_{ijk}T^i T^j T^k + \half K_{ijk} \tilde{f}^i T^j T^k - \bar{f}_i T^i + \bar{f}_0 - 2i f_0 \epsilon \;, \label{kwtcor}
\ea
where
\be
\bar{f}_i = f_i - f_0 K^{(2)}_i + 2 \tilde{f}^j K^{(1)}_{ij} \;,\;\; \bar{f}_0 = \tilde{f}_0 - \tilde{f}^i K^{(2)}_i \;.
\ee
Notice  that we can absorb the lower order corrections into a redefinition of the fluxes\footnote{This may be related to the combination $F_4 + F_0 B \w B$ which appears in massive IIA supergravity.}. We now go on to study moduli stabilisation using the corrected scenario of (\ref{kwtcor}).

\subsection{Solving the K\"ahler moduli F-terms}
\label{alphaftermsect}

The analysis of the K\"ahler moduli F-terms proceeds as in subsection \ref{SecFtermsim}, so we will be brief here and present only the results. 
The axions are fixed as in (\ref{bvev}), but the the K\"ahler moduli satisfy 
\be
9 \epsilon f_0^2 \kappa_{i} = \left( 3 \epsilon - 2\kappa \right) \left( K_{ijk} \tilde{f}^j \tilde{f}^k + 2 f_0 \bar{f}_i \right) \;. \label{kcorsol}
\ee
This is a non-trivial constraint  and gives a condition that  fixes the K\"ahler moduli at acceptable values. For the dilaton we find the following solution
\ba
s &=& \frac16 \frac{f_0}{h_0} \left( \kappa + 12 \epsilon \right)  + \frac{\tau^i}{2h_0f_0}  \left( K_{ijk}\tilde{f}^j\tilde{f}^k + 2\bar{f}_i f_0 \right) \nn \;, \\
\sigma &=& -\frac{1}{3h_0f_0^2} \left( K_{ijk}\tilde{f}^i\tilde{f}^j\tilde{f}^k +3 f_0 \tilde{f}^i\bar{f}_i \right) - \frac{\bar{f}_0}{h_0}\;. \label{scorsol}
\ea
The corrections therefore allow for a minimum in which the K\"ahler moduli and the dilaton are all fixed. Note that in (\ref{kcorsol})
the $\alpha'$ correction on the left hand side has to compete with the tree-level term on the right. However  we can still maintain a large vev 
for the $\tau^i$ fields by tuning the values of the fluxes. To see this more explicitly in a particular example, we can find a
solution to the implicit equation (\ref{kcorsol}) by focusing on the homogeneous case $\tau^i = \tau$, where we can define
$\kappa_i \equiv K_i \tau^2$, $\kappa \equiv K \tau^3$, $\tilde{f}^i \equiv \tilde{f}$ and $\bar{f}_i \equiv \bar{f} K_i$. We  then obtain  the solution
\ba
\tau &\simeq& \frac{-9\epsilon f_0^2}{2 K \left( \tilde{f}^2 + 2f_0 \bar{f} \right)} \;, \nn \\ \label{simskcor}
s &\simeq& \frac{f_0 K \tau^3}{6 h_0} \;,
\ea   
where we made the approximation $\kappa \gg \epsilon$. As expected, the vev of the K\"ahler moduli is proportional to the $\epsilon$ term, but we can still 
make this large by taking the flux $f_0$ large (whilst keeping $\tilde{f}$ and $\bar{f}$ small). Note that in this  limit $s$ is also large. However we should keep in mind that the value of $f_0$ is capped by the tadpoles constraints (\ref{tadpoles}) \footnote{The capping of the moduli vevs is typical behaviour when only the K\"ahler moduli or only the complex-structure moduli appear in the superpotential, while cases where both appear parametrically controlled solutions can be found \cite{Shelton:2005cf,DeWolfe:2005uu,Micu:2007rd}.}. 

\bigskip

Note that the IIA $\alpha'$ corrections we are considering  correspond  in the mirror  IIB picture  
to corrections away from the large complex-structure limit. Then the F-term equations we solved in this  IIA set-up are still 
equivalent to the ISD condition on the flux in IIB: in  Appendix \ref{AppendixA2} we show that this is indeed the case,
and that the ISD equations away from the large complex-structure limit are solved by (\ref{bvev}), (\ref{kcorsol}) and (\ref{scorsol}).

\subsection{The scalar potential}
\label{scalpotgenc}

It is interesting to see how this stabilisation mechanism  can be understood at the level of the scalar potential. 
This analysis will also be useful in understanding how the $\alpha'$ expansion is realised. 
  
To simplify the expressions we again focus on the case $b^i=\tilde{f}^i=0$. To calculate the scalar potential we can use the supergravity formula in terms of the K\"ahler potential and 
superpotential (\ref{kwtcor})\footnote{The corrected K\"ahler derivatives are given in (\ref{ecorkder}). Also $e^K=\frac{3R}{4s^4\kappa_3}$ and the following identity can be used to write it in the form (\ref{copt}):  $\frac{\kappa_3}{12} - \frac{3\kappa_3\epsilon}{8\kappa_{-3/2}} - \frac{81\kappa \epsilon^2}{32 \left(\kappa_{-3/2}\right)^2} + \frac{243\kappa^2\epsilon^2}{32\kappa_3\left(\kappa_{-3/2}\right)^2} = \frac{3}{4\kappa_3}\left( \frac{\kappa^2}{36} + \frac{2 \kappa \epsilon}{3} + 4\epsilon^2 + \frac{\kappa\kappa_{-6}}{12} \right)$
.}. We may also use the expression of (\ref{scapot}) with the $\epsilon$-corrected $N$ matrix of Appendix \ref{AppendixA2}. However, this will not account for the corrections to the first term in (\ref{scapot}).
The corrected scalar potential then reads 
\be
V = R\left[  \frac{3 h_0^2 r}{2s^2}\left(\frac{\kappa_{-\frac32}}{\kappa_3\kappa_{-6}}\right) + \frac{f_0^2\kappa_{3}}{12s^4} - \frac{f_0h_0}{s^3} -
 \frac{3f_0^2\epsilon\kappa_3}{8\,s^4\,\kappa_{-\frac32}}
-
\frac{L}{2s^4} \right] \;, \label{copt}
\ee 
where 
\ba
L &\equiv& 
 \left( \kappa^{ij} - \frac{3\tau^i\tau^j}{\kappa_3} \right) \hat{f}_i \hat{f}_j 
\;, \;\;
\hat{f}_i \equiv \bar{f}_i + \frac{9f_0\epsilon \kappa_i}{4\kappa_{-\frac32}} \;, \;\;
\kappa_x \equiv \kappa + x\epsilon \;.
\ea
We have already minimised with respect to the axion $\sigma$, which only appears in one positive-definite term and so just sets that term to zero and is fixed as in (\ref{scorsol}). The  equations we have to solve  are
\ba
\partial_s V &=& \frac{R}{s^5} \left[  -3rh_0^2 \left(\frac{\kappa_{-\frac32}}{\kappa_3\kappa_{-6}}\right)s^2 + 3f_0h_0s -\left( \frac{f_0^2\kappa_3}{3} -
\frac{3 f_0^2 \,\epsilon\,\kappa_3}{2\, \kappa_{-\frac32}}
- 2L \right)  \right]\,=\,0 \;,  \label{deras}\\
\partial_{\tau^i} V &=& \frac{R\kappa_i}{s^4} \left[ \frac{9rh_0^2s^2}{2\kappa_3\kappa_{-6}}\left( 1 - \frac{\kappa_{-\frac32}}{\kappa_3} -\frac{\kappa_{-\frac32}}{\kappa_{-6}} \right)  +\frac{f_0^2}{4}+
\frac{81}{16} \frac{f_0^2 \epsilon^2}{\kappa_{-\frac32}^2}
 \right]
-\frac{R}{2 s^4} 
\partial_{\tau^i}L\,=\,0 \label{deratau}
\ea
where 
\ba        
\partial_{\tau^l}\,L
&=&
- \left( 
\kappa^{i m} \kappa^{j n}\,
\kappa_{l m n}
+6 \frac{\delta_l^i \tau^j }{\kappa_{3}}
-9  \frac{\tau^i \tau^j}{\kappa_{3}^2}\,\kappa_l
\right)\,\hat{f}_i \hat{f}_j
\nonumber \\
&+& \frac{9 f_0 \epsilon}{\kappa_{-\frac32}}\, \left(\kappa^{ij} 
-3 \frac{\tau^i \tau^j}{\kappa_{3}}
\right)\,\hat{f}_i\,
\,\left( 
\kappa_{jl}-\frac32 \frac{\kappa_j \kappa_l}{\kappa_{-\frac32}}
\right)\,.\label{derEL}
\ea
In order  to  understand  how the stabilisation of the K\"ahler moduli occurs in this context, we plug  in the previous equations 
the solution for $s$ that we already found
\ba
s &=&
 \frac{f_0\kappa_{12}}{6h_0}   + \frac{\tau^i \,\bar{f}_i }{ h_0 }  
 = \frac{f_0}{6 h_0}\,\frac{\kappa_3 \kappa_{-6}}{\kappa_{-\frac32}}
+  \frac{\tau^i \,\hat{f}_i}{ h_0 }  \label{solps} 
\equiv
s_0+s_1\,\,,
\ea
where the last equality is such that $s_0$ and
$s_1$ are identified with the first and second terms respectively. 
We find, in the case
$r=2$,  the conditions
\ba
0 &=&  
\frac{\hat{f}_i\, \tau^i}{2} 
\,\left[   f_0 -
\frac{6\, h_0 \,\kappa_{-\frac32}\,s_1}
{
\kappa_3 \kappa_{-6}} 
\right] +
 \left( \kappa^{ij} - \frac{3\tau^i\tau^j}{\kappa_3} \right)\, \hat{f}_{i} \hat{f}_{j}\,, \label{confixk1}\\
0 &=& 
\frac{18\,(\hat{f}_i
\tau^i)\,h_0\, \kappa_l \, \left(s_1+2 s_0\right)}{ \kappa_3\kappa_{-6}} \,
\left( 1 - \frac{\kappa_{-\frac32}}{\kappa_3} -\frac{\kappa_{-\frac32}}{\kappa_{-6}} \right) - \partial_{\tau^l} L
\label{confixk2}
\ea
that indeed are solved by choosing  $\hat{f}_i=0$ which is equivalent to (\ref{kcorsol}). As expected from the no-scale structure, $V=0$ in the minimum. 
It is important to notice that, after expanding the expressions for $\hat{f}_i$, the conditions (\ref{confixk1}) and  (\ref{confixk2}) involve terms of order $\epsilon^2$,
and terms of this order are essential  to stabilise the K\"ahler moduli. Indeed, using (\ref{kcorsol}) we can write the solution for $s=\frac{f_0}{6h_0}\frac{\kappa_3\kappa_{-6}}{\kappa_{-\frac32}}$ which when substituted into (\ref{copt}) cancels the first four terms leaving only a term proportional to $L$.

 \bigskip
 
Note that the scalar potential (\ref{copt}) cannot be written in the form
(\ref{kachsca}),  and so avoids the no-go theorem of \cite{Hertzberg:2007wc}, which was derived at tree level in $\alpha'$ only. 
This will be important in section \ref{upsect}, where we will show it is possible  to uplift our minimum to a de Sitter vacuum using only $D6$ branes.

\subsection{The $\alpha'$ and $g_s$ expansions}
\label{algsexp}

Since the moduli stabilisation relies on $\alpha'$ corrections it is important to understand how the $\alpha'$ expansion can be kept under control. We now explore this in more detail. 
The K\"ahler moduli are fixed by competition between tree-level terms and $\alpha'$ corrections. In general, the $\alpha'$ expansion is essentially an expansion of the  four dimensional effective action in terms in the vevs of the K\"ahler moduli. As an example,  consider the second term of the potential (\ref{copt}) which contains a tree-level part and the relative $\epsilon \sim \alpha'^3$ correction
\be
\sim \frac{f_0^2}{\tau^3} + \frac{f_0^2\epsilon}{\tau^6} \;.
\ee
Here we see that the higher order term in $\alpha'$ is suppressed by a larger power of the K\"ahler moduli. This is always the case when the fluxes, and the powers of the dilaton, appear as a common factor in the  part of the potential that is being expanded.
  
Recall that fluxes are dimensionful quantities and, in particular, fluxes of different degree must be integrated over different degree cycles.
This implies that terms in the four-dimensional effective action associated with  different fluxes have a different dependence on the K\"ahler moduli.
Indeed this is the key property that allows K\"ahler moduli to be fixed perturbatively in IIA, while in IIB, where the fluxes are of the same degree (three-forms), the K\"ahler moduli dependence is universal and leads to a no-scale structure. The dependence of the power of the K\"ahler moduli on the degree of the flux implies that, for generic values of the fluxes, the $\alpha'$ expansion is not so clear-cut. For example, if we also include the tree-level term for $F_4$  we have
\be
\sim \frac{f_0^2}{\tau^3}  + \frac{f_0^2\epsilon}{\tau^6} + \frac{f_i^2}{\tau^7} \;. \label{alex2}
\ee    
Then for generic values of the fluxes $f_i$ and $f_0$, the $\alpha'$ corrections are lower order in the inverse vev expansion. Of course this just corresponds to the fact that the vev of the moduli is fixed by the fluxes and we must choose the fluxes so that their vev is large. So if the K\"ahler moduli were fixed by the competition between the tree-level terms in (\ref{alex2}) then we would have $\tau^2 \sim \frac{f_i}{f_0}$ and so to reach large values we must take $f_i \gg f_0$ which would mean that the third term in (\ref{alex2}) would still dominate the second.

The two cases above illustrate the two important properties of the $\alpha'$ expansion that we will use. The first is that for terms involving the same fluxes (and dilaton) factors, the $\alpha'$ expansion is exactly an expansion in the vev of the moduli and it is a valid expansion as long as the vev is large. In a way it is really this expansion that is the essence of the $\alpha'$ expansion and it is this one that we have well under control. The second point is that for terms with different degree fluxes the $\alpha'$ expansion is more complicated and is a property of the solution of the vev in terms of the fluxes. In our case we find that the fixing of the moduli requires competition between $\alpha'$ and tree-level terms in the sense of this latter expansion.   

Let us return to our solution in light of these considerations. In 
subsection \ref{alphaftermsect} we showed that once the dilaton is fixed at its minimum, the K\"ahler moduli are fixed by the requirement that $\hat{f}_i$ vanished. This can be seen as follows. Schematically, the potential (\ref{copt}) can be expanded as
\be
V \sim \frac{f_0^2}{\tau^3} + \frac{h_0^2}{\tau^6} + \frac{f_0h_0}{\tau^{\frac92}} + \frac{f_0^2\epsilon}{\tau^6} + \frac{\left(f_i + f_0 K^{(2)}_i \right)^2}{\tau^7} + \frac{\left(f_i + f_0 K^{(2)}_i \right)f_0\epsilon}{\tau^8} + \frac{f_0^2\epsilon^2}{\tau^9} \;+\;...
\ee 
where the ellipses denote higher order terms that are suppressed in the vacuum. These include terms that are lower order in the $\alpha'$ expansion such as $\frac{h_0^2\epsilon}{\tau^9}$.
Here we are interested in the powers of the K\"ahler moduli in relation to the $\alpha'$ expansion and so have factored out the appropriate powers from the definition of the four-dimensional dilaton superfield ($s \sim \tau^{\frac32} \tilde{s}$ where $\tilde{s}$ is independent of the K\"ahler moduli).  As we showed in section \ref{scalpotgenc}, minimising with respect to the dilaton nullifies the contribution from the first four terms. The remaining terms, which have the same dilaton factor in front, fix $\tau$ as in (\ref{kcorsol}).
  We therefore see that the K\"ahler moduli are fixed by competition between tree-level terms involving $f_i$ and higher order terms in $\alpha'$ up to  terms in $(\alpha')^6f_0^2$: this is the first non-vanishing term involving $f_0$ only\footnote{There is a subtlety here due to the lower order corrections $K^{(2)}_i$. These do induce lower order corrections in $f_0^2$. However they only appear in combination with the $f_i$ so that we are required to choose a cancellation between the two fluxes so that $\bar{f}_i$ is smaller than $f_0$. This is accounted for in (\ref{kcorsol}) where such choice is required to go to large K\"ahler vevs.}. Hence we keep only the first order (in $\alpha'$) term for each type of flux. 
  
It is important to note that the $\alpha'$ analysis we are doing is essentially an analysis of the mirror to the IIB ISD equations away from the large complex-structure limit. Since on the IIB side we expect the ISD equations to hold away from the large complex-structure limit, and since mirror symmetry should hold at all orders in $\alpha'$, we expect that our analysis captures the relevant corrections. 
 
 \bigskip
 
All of the analysis so far has been done at tree-level in terms of string loops. String loops are certainly expected to induce corrections in both the K\"ahler and complex-structure moduli \cite{Berg:2005ja}. However we will show in subsection \ref{csmoduli} that in these compactifications the ten-dimensional string coupling $g_s$ can be made exponentially small. The complex-structure `volume' $\Vol'$ on the other hand is exponentially large, and the two can compensate each other\footnote{For example, this allows the NS term in (\ref{scapot}), which is of lower order in string coupling, to compete with the other RR terms.}.  Correction terms where we do not expect such complex-structure moduli dependence, in particular terms such as $F^4_0$, we can safely neglect since the small string coupling will dominate any enhancement effects from flux values. The cases which involve the complex-structure moduli, such as KK and winding mode exchange between D6 branes, can introduce significant corrections and for those a more careful analysis must be made. We leave this to section \ref{csmoduli} and just state here that they are exponentially suppressed compared to the $\alpha'$ corrections. 

\subsection{Including the complex-structure moduli}
\label{csmoduli}

We can now proceed to the second stage of the moduli stabilisation. Having fixed the K\"ahler moduli by means of $\alpha'$ corrections, we can focus
on the stabilisation of the complex-structure moduli sector.  We include in our discussion non-perturbative effects, and the $\xi$ term in the complex-structure K\"ahler potential as derived in section \ref{csnoal}. The latter is associated with deviations from the large complex structure moduli limit and breaks the no-scale properties of the configuration. We will see that since this account for the IIB 
$\alpha'$ corrections, we are able to reproduce the mirror to the IIB LARGE volume model of \cite{Balasubramanian:2005zx} . 

We consider the superpotential  
\ba
W &=& W_0 + \sum_{\tilde{\lambda}} A_{\tilde{\lambda}}e^{-a_{\tilde{\lambda}}U_{\tilde{\lambda}}} \,\,\,\,.
\ea
Here we treat $W_0$ as a constant resulting from integrating out the K\"ahler moduli and dilaton.
We have separated the index range of the complex-structure superfields as $\lambda=\{b,\tilde{\lambda} \}$. This is in expectation that
one of the superfields, denoted by the index $b$, will take a much larger  vev than the other moduli. The new non-perturbative term in the complex-structure moduli 
can be attributed to gaugino condensation on D6 branes or E2 brane instantons\footnote{Note that the compatibility of these effects with an embedding of a visible chiral matter sector is non-trivial and may affect the moduli stabilisation scheme \cite{Blumenhagen:2007sm}.}.  
The non-perturbative term involving the quantity $U_b$ will be doubly exponentially suppressed, and so can be neglected. The associated
scalar potential can be written as
\ba
V &=& e^K \Big[  K^{T^i\bar{T}^j}F_{T^i}\bar{F}_{\bar{T}^j} + K^{S\bar{S}}F_S\bar{F}_{\bar{S}} + \left( K^{\bar{S}U_{\tilde{\lambda}}} F_{U_{\tilde{\lambda}}}\bar{F}_{\bar{S}} + c.c. \right) 
\nonumber \\
&+& 
K^{U_{\tilde{\lambda}}\bar{U}_{\tilde{\sigma}}}F_{U_{\tilde{\lambda}}}\bar{F}_{\bar{U}_{\tilde{\sigma}}} + \left( K^{U_b\bar{U}_b}K_{U_b}K_{\bar{U}_b} - 3 \right) |W|^2 \Big] \,\, ,
\ea 
where the F-terms have  their usual form $F_{S} = \partial_{S}W + (\partial_{S}K)W$. We will need the 
fact  that, for the vacuum we are interested in, the scaling of the different terms with respect to  $\Vol'$ go as
\ba
&& e^K \sim \Vol'^{-2} \hskip0.5cm ,\hskip0.5cm 
 F_{U_{\tilde{\lambda}}} \sim F_{S} \sim F_{T} \sim K^{\bar{S}U_{\tilde{\lambda}}}  \sim \left(K^{U_b\bar{U}_b}K_{U_b}K_{\bar{U}_b} - 3\right) \sim \Vol'^{-1}
\,\,, \nonumber \\
   && K^{T\bar{T}} \sim K^{S\bar{S}} \sim W \sim 1 \hskip0.5cm,\hskip0.5cm  K^{\bar{U}_{\tilde{\lambda}}U_{\tilde{\sigma}}} \sim \Vol' \;. \label{scalings}
\ea
Accepting this  for the moment, and keeping only terms  up to order $\Vol'^{-3}$,  we get the potential
\ba
V \,=\, \frac{1}{32s\Vol} &\Big[&\frac{4}{\Vol'} \left( -d_{\tilde{\lambda}\tilde{\sigma}\tilde{\rho}} q^{\tilde{\rho}}\right) A_{\tilde{\lambda}} a_{\tilde{\lambda}} e^{-a_{\tilde{\lambda}} U_{\tilde{\lambda}}}  A_{\tilde{\sigma}} a_{\tilde{\sigma}} e^{-a_{\tilde{\sigma}} U_{\tilde{\sigma}}}  - \frac{2u_{\tilde{\lambda}}}{\Vol'^2} \left( A_{\tilde{\lambda}} a_{\tilde{\lambda}} e^{-a_{\tilde{\lambda}} U_{\tilde{\lambda}}}\bar{W}_0 + c.c. \right) \nonumber \\&+& \frac{3\xi'|W_0|^2}{4\Vol'^3}  \,\,\Big] \;.
\ea
We can further  simplify the form of the potential,  requiring  that the CY has a "mirror Swiss-cheese" form, so that $\Vol'$ reads
\be
\Vol' = \frac16 d_{\lambda\sigma\rho}\,q^{\lambda}q^{\sigma}q^{\rho} = \alpha \left( u_b^{3/2} - h^{\tilde{\lambda}} u_{\tilde{\lambda}}^{3/2} \right) \;. \label{cheese}
\ee
In that case we can write 
\be
 -d_{\tilde{\lambda}\tilde{\sigma}\tilde{\rho}} \,q^{\tilde{\rho}} \simeq \frac{2\,\Vol' \,u_{\tilde{\lambda}}^{1/2}\, \delta_{\tilde{\lambda}\tilde{\sigma}} }{3\,\alpha \,h^{\tilde{\lambda}}} \;.
\ee
This gives the scalar potential
\be
V = \frac{1}{32s\Vol} \left[  \frac{8 u_{\tilde{\lambda}}^{1/2} }{3\Vol' \alpha h^{\tilde{\lambda}}} \left| A_{\tilde{\lambda}} a_{\tilde{\lambda}} \right|^2
 e^{-2a_{\tilde{\lambda}} u_{\tilde{\lambda}}} - \frac{4u_{\tilde{\lambda}}}{\Vol'^2}\left| A_{\tilde{\lambda}} a_{\tilde{\lambda}} \right| |W_0|  e^{-a_{\tilde{\lambda}} u_{\tilde{\lambda}}} 
 + \frac{3\xi'|W_0|^2}{4\Vol'^3} \right] \;, \label{scapot2}
\ee
where we have fixed the axions $\nu_{\tilde{\lambda}}$,  as they adjust to make the sign of the second term of (\ref{scapot2}) negative\footnote{For the case of a single axion this is true. As pointed out in \cite{Blumenhagen:2007sm},  the case of multiple axions is not so clear and the interplay between the phases could lead to modifications of the scenario.}.
This potential was shown in \cite{Balasubramanian:2005zx}  to admit a non-supersymmetric AdS  minimum with all  moduli fixed, 
where  $\Vol' \sim u_b^{3/2}$ is exponentially large where $\left( \mathrm{ln}\; \Vol' \right) \sim a_{\tilde{\lambda}} u_{\tilde{\lambda}}$. 

At the minimum,  the scaling properties  with respect to $\Vol'$ (\ref{scalings}) follow simply from the form of the K\"ahler potential, apart from the expressions for the dilaton and K\"ahler moduli F-terms. To calculate their scaling behaviour we consider a small perturbation around the point $F_S = F_{T^i} =0$. Let us consider just the dilaton for simplicity and write it as $s = s_0 + \delta_s$. Then we can expand
\ba
K^{\bar{S}S}\bar{F}_{S}F_S = M \delta_s^2 \hskip0.5cm&,& \hskip0.5cm  |W| = |W_0| + |W_1| \delta_s \;,\;\;
\nonumber \\ 
 K = K_0 + K_1 \delta_s \hskip0.5cm&,& \hskip0.5cm
V = N e^{K_1\delta_s} \left[ M \delta_s^2 - \gamma \delta_s - \alpha \right] \;.
\ea 
Here $|W_0|$ and $K_0$ denote the superpotential and K\"ahler potential evaluated at $s=s_0$, and $W_1$ and $K_1$ are the first terms in the expansion. 
$M$ is a (positive) constant of order one, whilst $\gamma$ and $\alpha$ are constants of order $\Vol'^{-1}$. Minimising this with respect to $\delta_s$ provides  $\delta_s \sim \Vol'^{-1}$. This gives the correct scaling (\ref{scalings}). The same analysis also holds for the K\"ahler moduli. The fact that the dilaton and K\"ahler moduli are expected to be fixed  very close to values corresponding
to  vanishing F-terms was already pointed out (in the IIB mirror) in \cite{Balasubramanian:2005zx}. There it was argued that moving away from vanishing F-terms would give a positive contribution to the potential, which overwhelms the other terms and so must constitute an increase in energy. We have confirmed this argument in showing that although the moduli  do
actually move from their supersymmetric values, the resulting contribution in the potential is suppressed with respect to the other terms.

It is interesting to evaluate the value of the ten-dimensional dilaton, corresponding
to the string coupling, in our  vacuum.  We get
\be
g_s^{-1} = e^{-\hat{\phi}} \simeq \sqrt{2} s^{\quarter} \Vol^{-\half} \Vol'^{\half} \;.
\label{vevgs}
\ee
Since $\Vol'$ is exponentially large, we find that we are at exponentially WEAK string coupling. The relation of this set-up to its LARGE volume mirror can be understood from the fact that T-duality acts non-trivially on the dilaton mixing it with metric components. We discuss some consequences of (\ref{vevgs}) in section \ref{Secdiscuss} where we argue that it can naturally lead to TeV scale supersymmetry breaking along with an intermediate string scale. Indeed the value of the string coupling can be considered the key feature of these compactifications.

\bigskip

Finally, we return to the issue of string loop corrections. In order for the above scenario to hold we require that these are suppressed with respect to the $\alpha'$ corrections (so that the K\"ahler moduli stabilisation is valid) and also with respect to the $\xi$ correction (so that the complex-structure stabilisation is valid). Here we simply outline an argument following \cite{Berg:2007wt,Cicoli:2007xp} as to why we expect this to be the case.

Since string loop corrections have only been explicitly computed in torodial models (see \cite{Angelantonj:2002ct,Berg:2005ja} for example), we have no direct calculation of these corrections. However it is still possible to guess the form of the corrections using the torodial result as in \cite{Berg:2007wt}. Further it was argued in \cite{Cicoli:2007xp} that the corrections can also be understood from a four-dimensional point of view as corrections suppressed by the gauge-coupling of the brane responsible for them. 
In IIB it was argued that these corrections are always suppressed by the CY volume due to the Weyl rescaling from the string to the Einstein frame. Then there are extra factors of powers of the size of the cycle wrapped by the brane corresponding to the masses of the exchanged KK or winding modes. Since winding modes become heavier for larger cycles this will just lead to extra suppression of the corrections in the cycle volumes. Therefore the leading corrections in that sense are the KK modes.  We are particularly interested in corrections to the LARGE volume models in IIB (although the following analysis, at least in terms of the relative sizes of the $\alpha'$ and $g_s$ corrections also holds for KKLT-like scenarios). In that case it was argued in \cite{Berg:2007wt} that they take the form
\be
\delta K_{g_s} \sim \frac{\sqrt{u_s}}{s\Vol'} + \frac{\sqrt{u_b}}{s\Vol'} \;. \label{strlpcor}
\ee
Here the corrections are given in terms of the corrections to the K\"ahler potential. We have translated the IIB result to our IIA language using the dictionary as in section \ref{cycomwithout}. 
We want to consider the case where $\Vol'$ is exponentially large and have restricted to a toy case where there are two complex-structure moduli, one exponentially large $u_b$ and one small $u_s$.  On the IIA side, we can understand the form of the corrections as follows. The suppression in $\Vol'$ comes from the extra factor of $g_s^2$ in the loop corrections. The factors in the numerators are the masses of the KK states.  We see that such corrections are subdominant to the $\alpha'$ corrections since they are suppressed by (a positive power of) $\Vol'$ whereas the $\alpha'$ corrections are only suppressed by $\kappa$.  The second term in (\ref{strlpcor}) leads to a correction that is dominant over the $\xi$ correction in the K\"ahler potential. However a cancellation in the scalar potential means that it is subdominant in the potential \cite{Berg:2007wt,Cicoli:2007xp}. This result was interpreted from a four-dimensional point of view in \cite{Cicoli:2007xp} where it was related to a cancellation in the Coleman-Weinberg potential. 

\section{Uplifting with D6 branes}
\label{upsect}

In the previous sections we have argued that the scalar potential admits a stable, non supersymmetric AdS minimum. In this section we attempt to uplift this AdS vacuum to a de Sitter one.
This is particularly interesting when the uplift is performed using D6 branes, which is probably the best understood possibility, since in this case not including  $\alpha'$ corrections implies that
it is not possible to obtain a de Sitter minimum \cite{Hertzberg:2007wc}~\footnote{An attractive alternative to D6 branes would be NS5 branes as considered in \cite{Looyestijn:2008pg}, or even a direct compactification to de Sitter \cite{Silverstein:2007ac},  although the latter requires manifolds that are not CY and breaks supersymmetry at a high scale.}. We show that for our vacua, thanks to the $\alpha'$ corrections, this conclusion  does not hold and de Sitter minima can be constructed, provided that the usual tuning requirements  on  the uplifting sector are imposed. 

\subsection{The uplifting mechanism}

The mechanism that we use is the introduction of D6 branes that form non-trivial angles with the O6 planes. 
This is in some cases dual \cite{Villadoro:2006ia} to the magnetised D7-branes uplifts in IIB \cite{Burgess:2003ic} 
and also the $\bar{D}3$ ones \cite{Kachru:2003aw}. Throughout our analysis we, for simplicity, neglect any world-volume fields. 
The vevs of these fields could play an important role in the uplifting procedure by compensating the contribution to the effective potential \cite{Burgess:2003ic}. 
However,  as shown for example in \cite{Achucarro:2006zf,Cremades:2007ig},  this need
not always to be the case,  and in some situations they may actually help with the fine-tuning needed for the uplift. Either way this  is a model dependent issue that
we leave  for  future work.     

\smallskip

We start by recalling that a calibration $\omega$ is a 
form  such that  its pullback gives  the world-volume 
of  a D6-brane/$O6$-plane wrapping a cycle
 \cite{Becker:1995kb}. 
  Schematically  we can write 
\be
\phi^* \omega = \sqrt{\mathrm{det\;}\left( \phi^*\left( g+B_2 \right) + F_2 \right)} d^p\sigma \;,
\ee
where $\phi^*$ is the pull back of the space-time fields to the brane world-volume,
and $F_2$ is the world-volume gauge field strength.  
We want to consider a D6/$O6$ plane wrapping a 3-cycle, then we have  \cite{Grimm:2005fa}
\be
\omega_{O6} = \re{\left( \sqrt{2} e^{\half \left(K^T-K^{cs}\right)} e^{-i\theta} \Omega \right)} = e^{\hat{\phi}} 2\re{(C\Omega)} \;. 
\ee  
We also have to impose the constraints
\be
\phi^* \left( J + iB \right) + 2 \pi i \alpha ' F_2 = 0 \;,
\ee
which imply that  $\phi^*J=0$ (the second calibration condition),
  and that we can not have any H-flux without a local source
  (the Freed-Witten anomaly cancellation condition \cite{Freed:1999vc}, \cite{LoaizaBrito:2006se}).
   This means that we should not wrap a D6 brane on $\beta^0$. Once we introduce charges into our setup we must satisfy the tadpoles constraints (\ref{tadpoles}). Consider now a calibrated D6 brane\footnote{We consider just a single D6 brane and a single orientifold. The case with multiple branes/orientifolds, at the level of our analysis, is just given by the appropriate choice of wrapping numbers for their cycles. } wrapping a generic three-cycle $\pi$. Its world-volume action then reads
\be
S_{wv} = \mu_6 \int_{\pi} \re{( 2e^{i\theta'} C \Omega)} \;, 
\ee
where we have included an angle $\theta'$,
 since it needs to  be calibrated with respect to the same form as the orientifold,
  up to a phase. In our conventions, where $2\pi\sqrt{\alpha'}=1$, we have $\mu_6=\frac{1}{2\kappa_{10}}$. This is the same factor appearing
    outside the bulk action and so we can factorise it out and take $\mu_6=1$ henceforth.
As shown in \cite{Becker:1995kb} we should adjust the angle $\theta'$ to maximise the action so that  we end up with 
\be
S_{wv} = \left| \int_{\pi} 2C\Omega \right| \;.
\ee
We then define 
\be
\tilde{\Omega}_{\pi} \equiv  \int_{\pi} 2C\Omega \;,
\ee
so
that the resulting (string frame) scalar potential from reducing the action
 is 
\ba
V^s_6 &=& \left(  |\tilde{\Omega}_{\pi}| + |\tilde{\Omega}_{\pi'}| - 4 |\tilde{\Omega}_{\omega}| \right) \; \nn \\
 &=& 2\left(  |\tilde{\Omega}_{\pi}| - \re{(\tilde{\Omega}_{\pi})} \right) + 2\re{(\tilde{\Omega}_{\pi})} - 4 \re{(\tilde{\Omega}_{\omega})} \nn \\
 &=& 2\left(  |\tilde{\Omega}_{\pi}| - \re{(\tilde{\Omega}_{\pi})} \right)  - f_0 h_0 \re{(\tilde{\Omega}_{0})} \nn \\
 &=&   2\left(  |\tilde{\Omega}_{\pi}| - \re{(\tilde{\Omega}_{\pi})} \right)  + f_0 \im{(W^Q)} \equiv V^s_D + V^s_F \label{vgendo} \;,
\ea
Here we included the contribution from the orientifold mirror of the D6 brane wrapped on the mirror cycle $\pi'$. The orientifolds are taken to wrap the cycle $\omega$. In 
passing from the first to the second line
 we have used the fact that since the orientifold is calibrated with respect to
  $C\Omega$, it satisfies $ |\tilde{\Omega}_{\omega}| =  \re{(\tilde{\Omega}_{\omega})}$. We also used the fact that the orientifold constraints (\ref{orcon}) imply that the real part of $C\Omega$ is proportional to the orientifold odd forms and the imaginary part to the even forms. Going to the mirror three-cycle just changes the phase of $\tilde{\Omega}_{\pi}$ and leaves $|\tilde{\Omega}_{\pi'}|=|\tilde{\Omega}_{\pi}|$. Finally we use the tadpole constraints to eliminate the local sources for the fluxes,
   and also the fact that $\im{(W^Q)} = \int_{CY}2 \re{(C\Omega)}\w H_3 =-h_0\re{(\tilde{\Omega}_{0})}$. We therefore recover the local contribution we have been using in sections \ref{scakpot} and \ref{scalpotgenc}, plus an additional D-term contribution. 

\smallskip

We now go on to analyse $V_D$. We want to work in the Einstein frame, so we must rescale by the Weyl rescaling 
 factor $e^{4D}$. We can write the potential as in \cite{Villadoro:2006ia}
\be
V_D = e^{4D} \frac{\im{(\tilde{\Omega}_{\pi})}^2}{\re{(\tilde{\Omega}_{\pi})}} \frac{2}{\sqrt{1 + \left(  \frac{\im{(\tilde{\Omega}_{\pi})}}{\re{(\tilde{\Omega}_{\pi})}} \right)^2} +1}
\ee
In \cite{Villadoro:2006ia} it was shown that only in the case $ \left|  \frac{\im{(\tilde{\Omega}_{\pi})}}{\re{(\tilde{\Omega}_{\pi})}} \right| << 1$ this
 can  be interpreted as a D-term, otherwise supersymmetry 
 would be
  broken non-linearly. Such cases,
  for example,
   correspond to the mirrors of $\bar{D}3$ in IIB.
    For the rest of this analysis we will consider only the cases where the limit is satisfied and this will place some mild constraints on our configurations. Then we have
\be
V_D = e^{4D} \frac{\im{(\tilde{\Omega}_{\pi})}^2}{\re{(\tilde{\Omega}_{\pi})}} \;.
\ee 
To evaluate this we expand the cycle
\be
\pi = e_0 \alpha_0 + m_{\lambda} \alpha_{\lambda} + e^{\lambda} \beta^{\lambda} \;,
\ee
which gives 
\be
V_D =  \frac{\left( m_{\lambda}  q^{\lambda} \right)^2}{4\left(\Vol' + \half \xi' \right)^2\left( e_0 s + e^{\lambda}  u_{\lambda} \right)} \label{dterms}
\ee
This will constitute
 our uplifting term. We note that since $q^{\lambda}$ and $u_{\lambda}$ have a factor of $s$ inside them, the term scales like $s^{-3}$ which matches the scaling expected from a $D6$ term. 

\subsection{Examples}

\begin{figure}
\includegraphics[width=.4\textwidth]{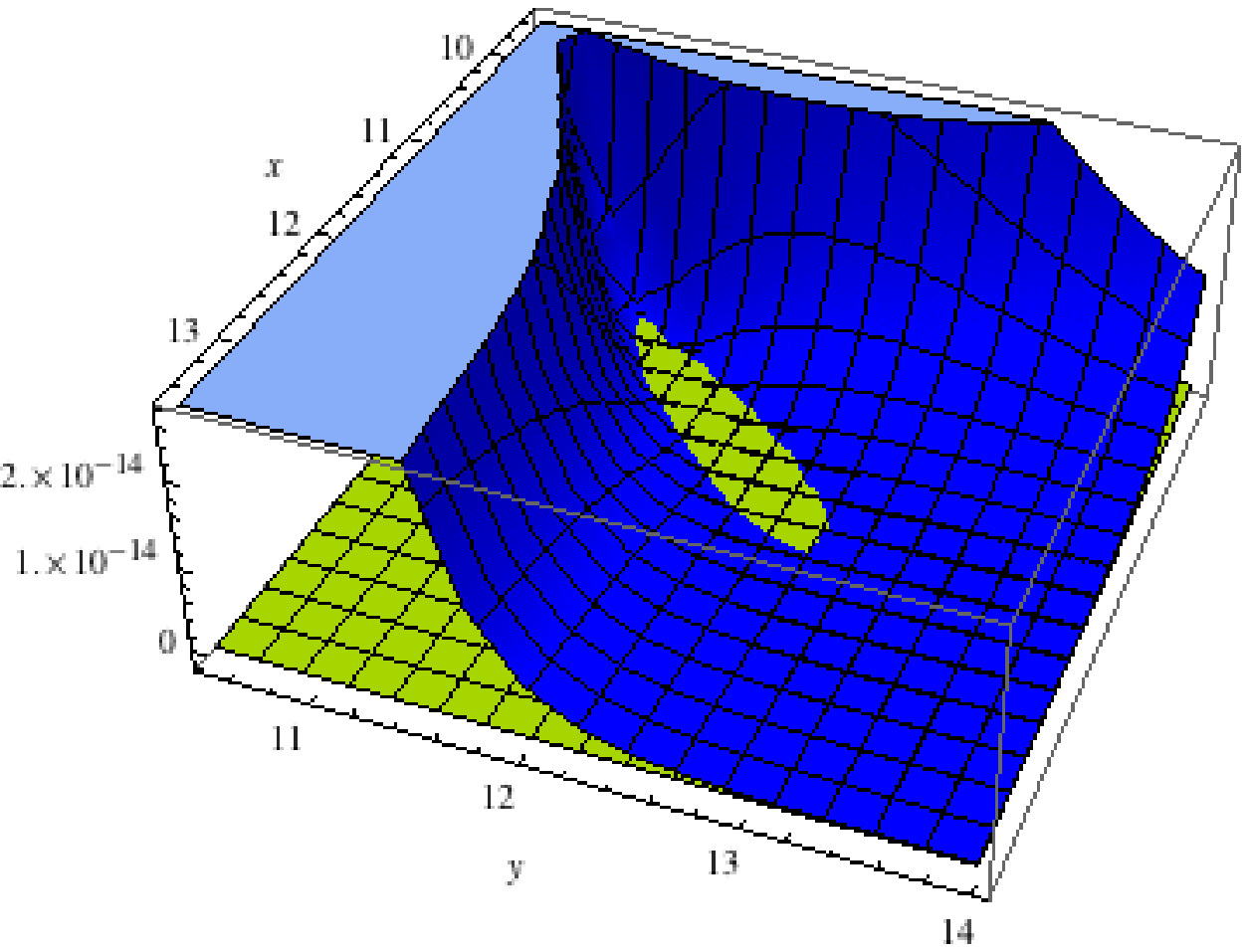}
\hspace{.3in}
\includegraphics[width=.4\textwidth]{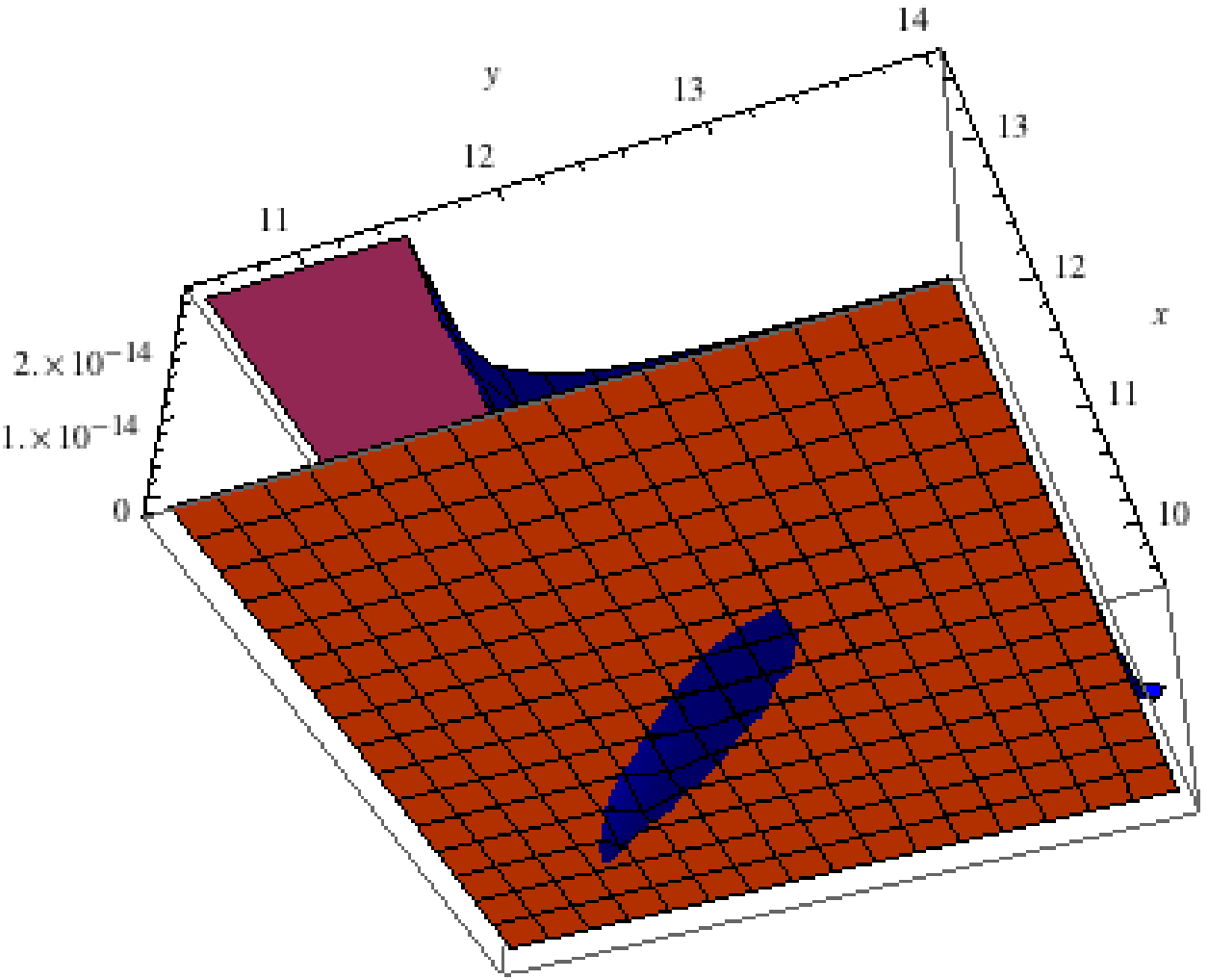}
\caption{The AdS minimum from above and below with $x=u_s$ and $y=\mathrm{ln}\;\Vol'$. The yellow/orange plane is at $V=-2\times 10^{-16}$.}
\label{fig:adsmin} 
\end{figure}

We now wish to study if this type of term in the scalar potential can be used to uplift the AdS vacua. This will generically depend on a large number of parameters,
 and so for manageability purposes we make some simplifications. We consider a simple model with only two complex-structure moduli. We can take this to be the mirror of $P_{[1,1,1,6,9]}$ with $h^{(2,1)}=2$ and $h^{(1,1)}=272$, in which case we have the two moduli $u_b$ and $u_{\tilde{\lambda}}=u_s$. In the notation of (\ref{cheese}) we have $\alpha=\frac{1}{9\sqrt{2}}$, $h^s=1$ and $\xi \simeq \frac43$. We take the gauge group such that $a_s=1$.  We also have the four continuous parameters $A_s$, $W_0$, $s_0$ and $\Vol_0$ which are in principle tunable using the $546$ RR fluxes at our disposal. To make the potential neater we take $|W_0|=3\sqrt{2}|A_s|$ and $|A_s|= s_0^{\frac32}\frac{\sqrt{2}}{9 (10)^{\frac32}}$. With this we can write the scalar potential as
\be
V = \frac{\sqrt{u_s}e^{-2u_s}}{\Vol'} - \frac{2u_s e^{-u_s} }{\Vol'^2} + \frac{10^{\frac32}}{\Vol'^3} + \frac{10^3\Vol_0}{s_0^2\Vol'^2}\frac{u_s}{e_0 s_0 + e^b u_b + e^s u_s} \;.
\ee
Here we neglected an overall constant multiplicative factor which does not alter the position or nature of the minimum. We also set $m_b=0$ and $m_s=1$ \footnote{The case $m_b \neq 0$ will washout the minimum unless also $e^b \neq 0$ in which case it reduces to case 2 in our analysis. }. Without the uplift term this potential has an AdS minimum at $u_s \sim 11$ and $\Vol' \sim e^{12} \sim 10^5$, plotted in Figure \ref{fig:adsmin}. The minimum can be seen as the isolated region below the plane at $V=-10^{-20}$. Including the uplift term there are three possible scenarios according to which of $\{e_0,e^b,e^s\}$ is non-vanishing\footnote{A combination of parameters not vanishing does not lead to any qualitatively new scenarios.}. They all share the feature that the value of $\Vol_0$ and $s_0$ must be tuned so that the uplift term does not wash out the minimum. 

\begin{figure}
\centering
\subfigure[The dS minimum, with uplift term $V_D=\alpha u_s\Vol'^{-\frac83}$ with $\alpha=1.0323\times 10^{-2}$, from above and below with $x=u_s$ and $y=\mathrm{ln}\;\Vol'$. The yellow plane is at $V=10^{-20}$ and the red plane is at $0$.]{
          \label{fig:dsmin1}
            \includegraphics[width=.4\textwidth]{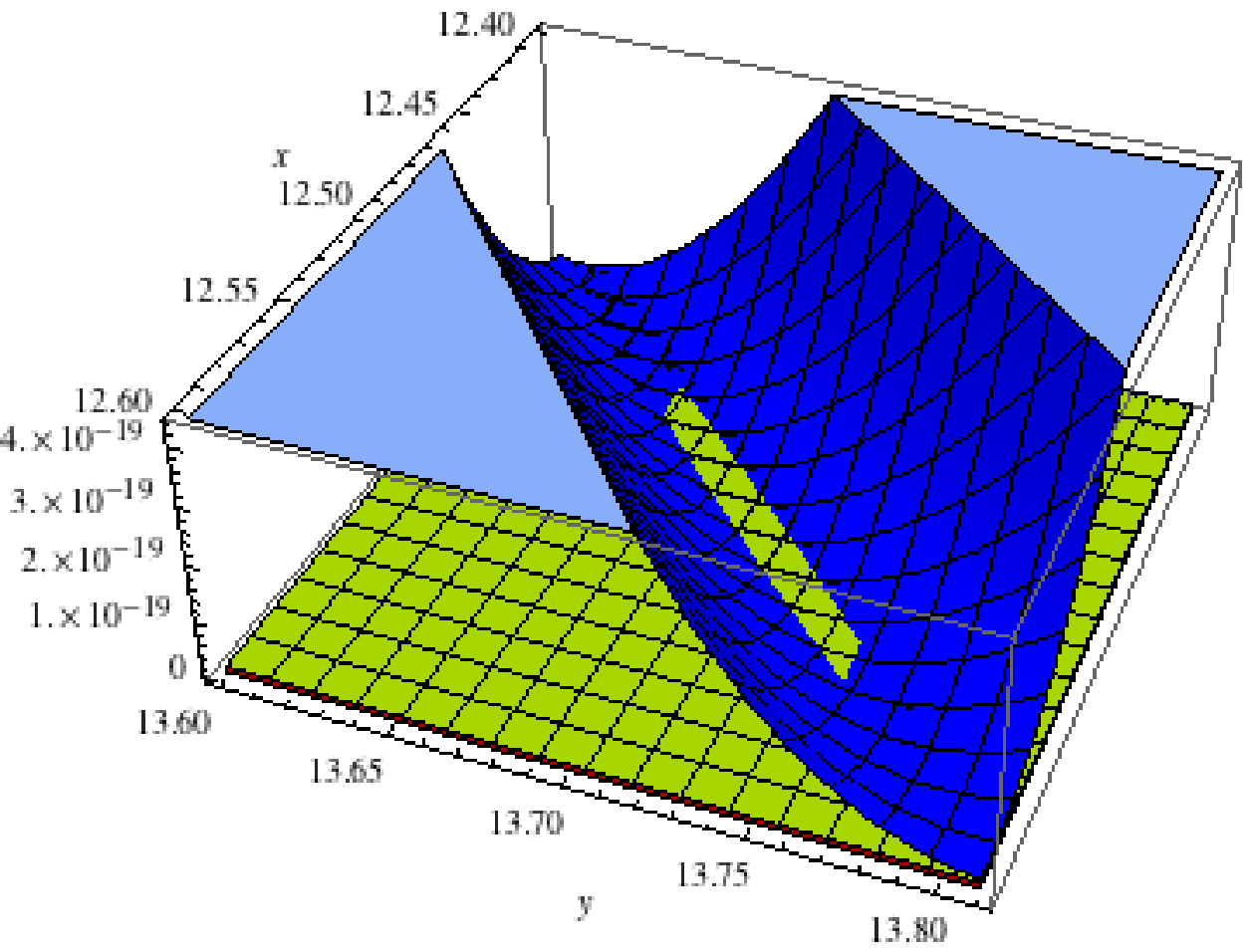}
\hspace{.3in}
 \includegraphics[width=.4\textwidth]{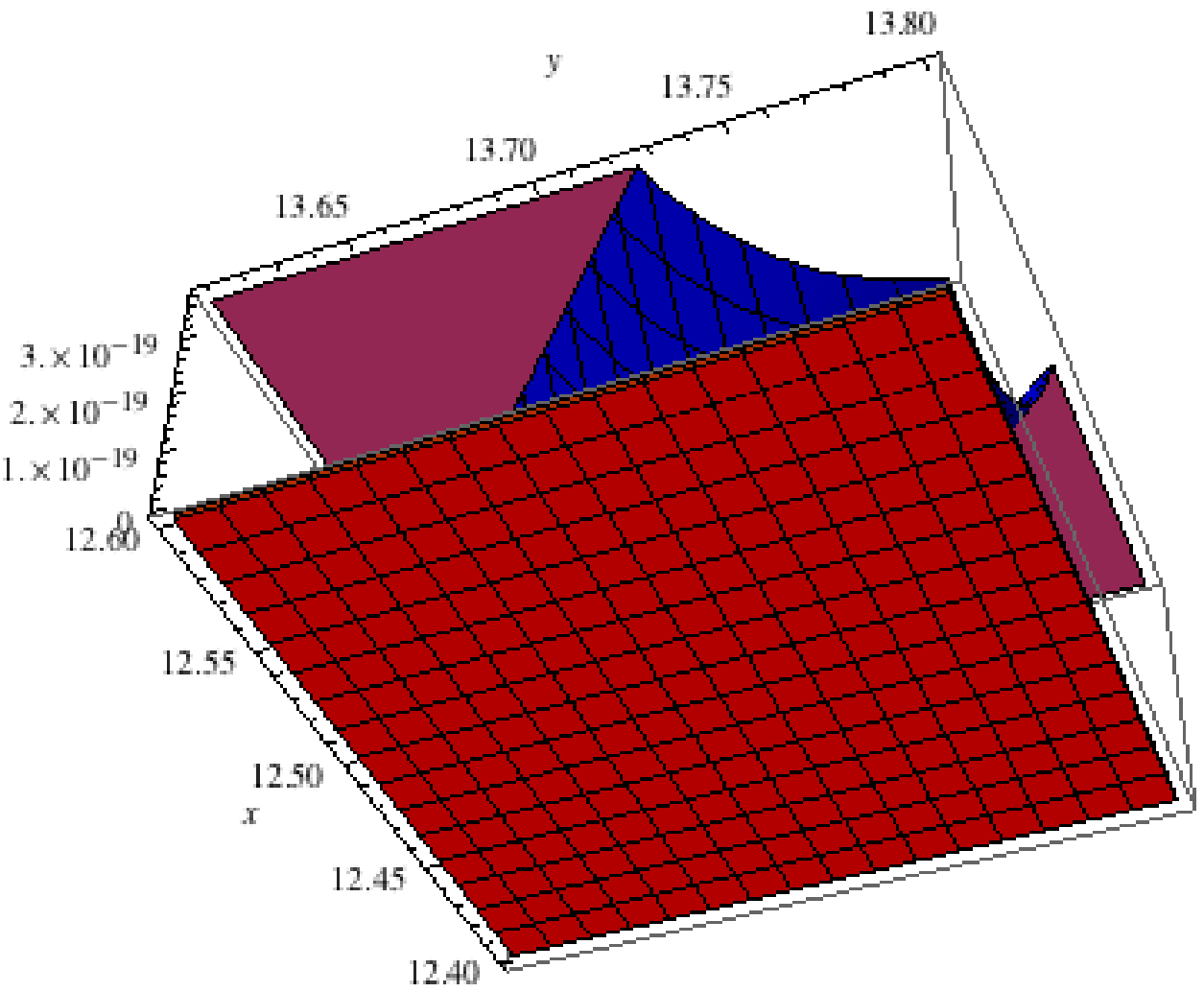}
 }
  \\ \vspace{.1in}
\subfigure[The dS minimum, with uplift term $V_D=\alpha\Vol'^{-2}$ with $\alpha=2.53504\times 10^{-5}$, from above and below with $x=u_s$ and $y=\mathrm{ln}\;\Vol'$.  The yellow plane is at $V=10^{-20}$ and the red plane is at $0$.]{
          \label{fig:dsmin2} \includegraphics[width=.4\textwidth]{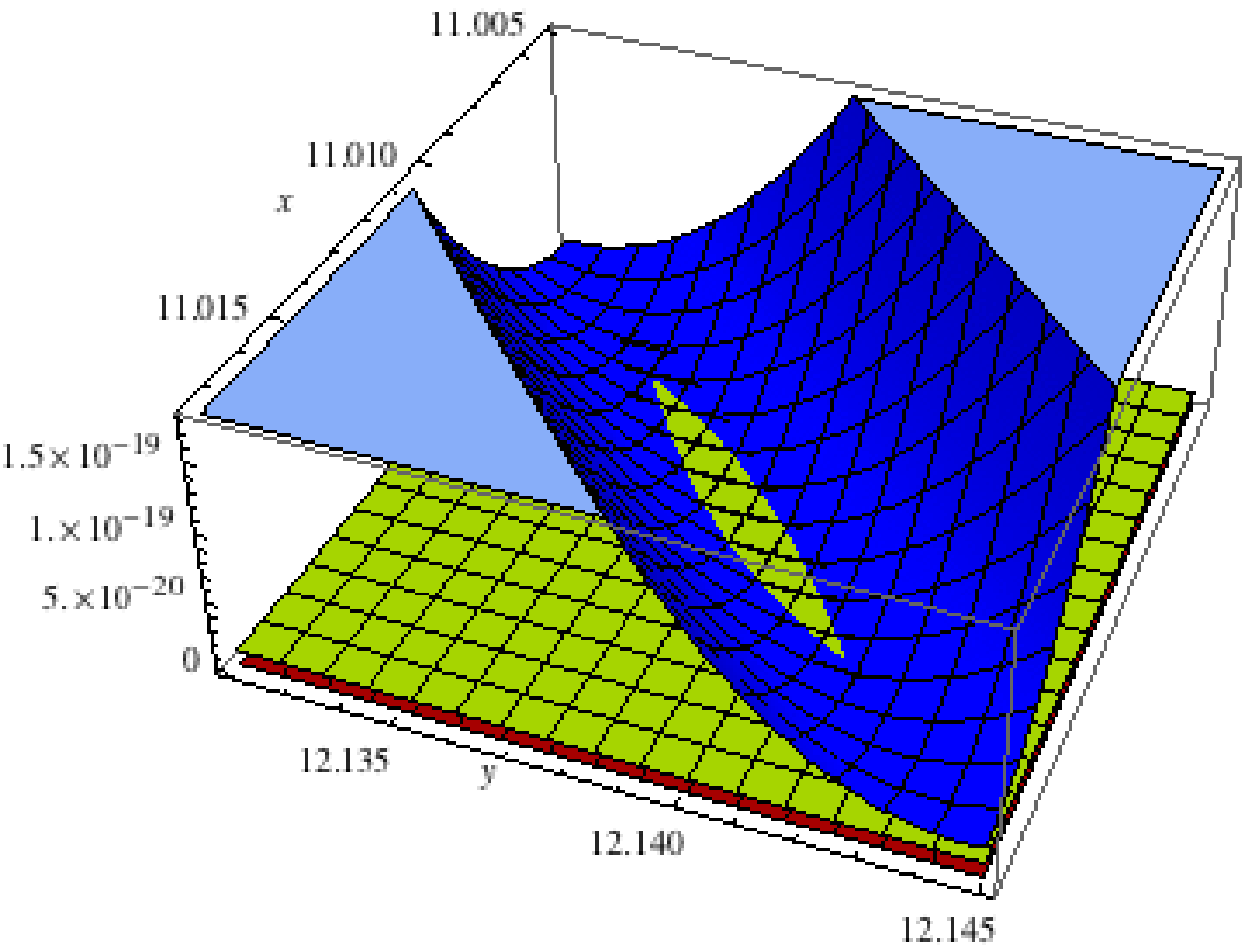}
\hspace{.3in} \includegraphics[width=.4\textwidth]{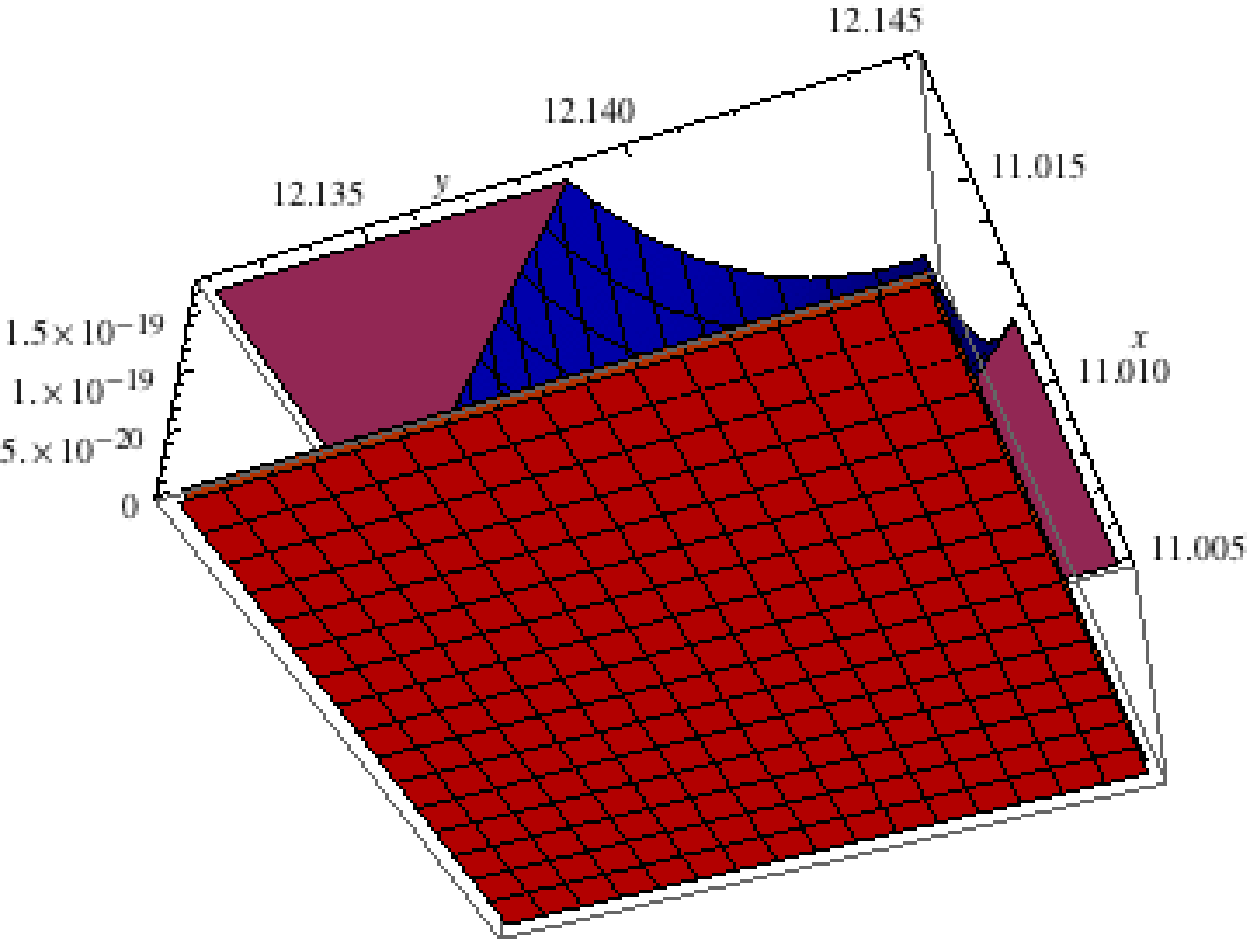}
} \\ \vspace{.1in}
\subfigure[The dS minimum, with uplift term $V_D=\alpha u_s\Vol'^{-2}$ with $\alpha=2.31147\times 10^{-6}$, from above and below with $x=u_s$ and $y=\mathrm{ln}\;\Vol'$.  The yellow plane is at $V=10^{-20}$ and the red plane is at $0$.]{
          \label{fig:dsmin3} \includegraphics[width=.4\textwidth]{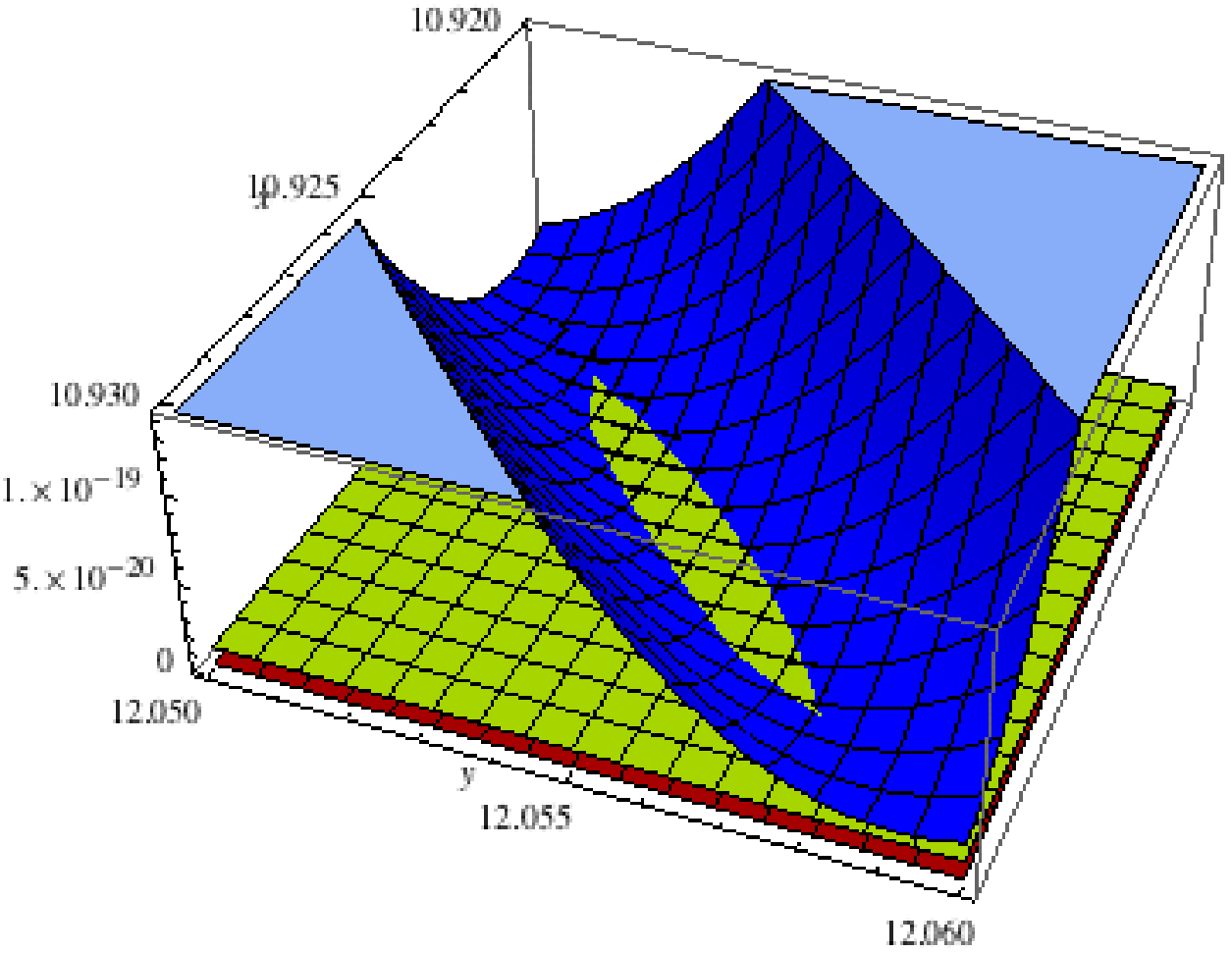}
\hspace{.3in} \includegraphics[width=.4\textwidth]{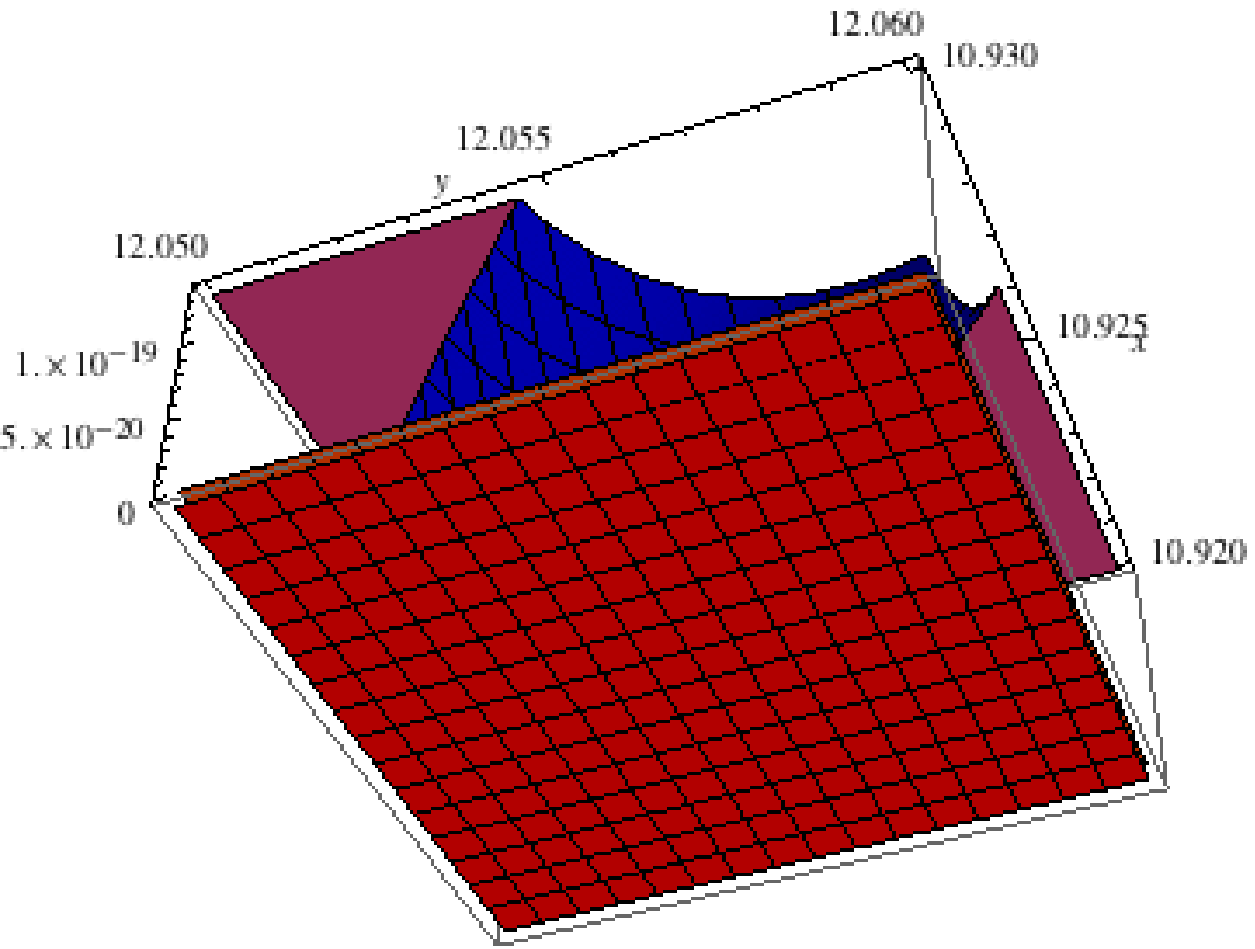}
} \\ \vspace{.1in}
\end{figure}

\medskip
\begin{itemize}
\item
\textbf{Case 1: $e^b \neq 0$}

\medskip 
\noindent
In this case  the uplift term is
\be
V_1 =  \frac{10^3\Vol_0}{s_0^2e^b}\frac{u_s}{\Vol'^{\frac83}} \equiv \alpha  \frac{u_s}{\Vol'^{\frac83}}\;.
\ee
This is plotted in Figure \ref{fig:dsmin1}, where a de Sitter minimum can be seen as the region below the plane at $V=10^{-20}$, but above $0$. The value of $\alpha$ required is $1.0323\times10^{-2}$. The tuning in this number is slightly exaggerated since a de Sitter minimum at a value above $V=10^{-20}$ would still exist for slightly different values. Nonetheless the tuning in $\alpha$ is a measure of the amount of tuning needed for such an uplift to work. To generate the order of magnitude for $\alpha$, we require a large enough vev for the dilaton and/or large wrapping number (though the exact values will differ according to more general values of the parameters $W_0$ etc). 
We can also check that supersymmetry is broken softly since 
\be
 \left|  \frac{\im{(\tilde{\Omega}_{\pi})}}{\re{(\tilde{\Omega}_{\pi})}} \right|  \sim \frac{s_0^{\half} q^s}{e^b \Vol'^{\frac23}} << 1 \;. 
\ee   

\medskip
\item
\textbf{Case 2: $e^s \neq 0$}

\medskip 
\noindent
In this case the uplift term is 
\be
V_2 = \frac{10^3\Vol_0}{s_0^2e^s}\frac{1}{\Vol'^{2}}   \;.
\ee
The de Sitter minimum resulting from this is shown in
Figure \ref{fig:dsmin2}. It can be seen that the tuning needed in this case is a few orders of magnitude larger. This is
because the discrepancy between the powers of $\Vol'$ in F-terms and D-terms is larger, and must be compensated by hand. 
This means that quite a large wrapping number and/or a large vev  for $s_0$. The soft-breaking parameter reads
\be
 \left|  \frac{\im{(\tilde{\Omega}_{\pi})}}{\re{(\tilde{\Omega}_{\pi})}} \right|  \sim \frac{s_0^{\half}}{e^s q^s} \sim   \frac{s_0^{\half}}{e^s} \;.
\ee
This can be made small consistently with $\alpha$ being small, by a large enough wrapping number (e.g. $e^s \sim s_0$). 

\medskip

\item 
\textbf{Case 3: $e_0 \neq 0$}

\medskip 
\noindent
In this case
\be
V_3 = \frac{10^3\Vol_0}{s_0^3e_0} \frac{u_s}{\Vol'^{2}} \;.
\ee
The de Sitter minimum resulting from this is shown in Figure 
\ref{fig:dsmin3}. The tuning is roughly of the same order as Case 2, but the suppression is enhanced
 by an extra power of the dilaton vev. The soft-breaking parameter reads
\be
 \left|  \frac{\im{(\tilde{\Omega}_{\pi})}}{\re{(\tilde{\Omega}_{\pi})}} \right|  \sim \frac{1}{e_0 s_0^{\half}} \;,
\ee
which can easily be made small. 

\end{itemize}
\bigskip
So  
we have seen that, given the usual caveats about tuning conditions, it is possible to find stable de Sitter vacua using only D6 branes in our setup. 
For complex-structure `volumes'  much larger than the ones we  considered,  for example $\Vol' \sim 10^{15}$, the tuning becomes more drastic and it seems that Cases 2 and 3 would struggle to suppress the uplift term enough. Case 1, which only has to make up a factor $\Vol'^{\frac13}$, could in principle allow for a minimum for some values of the parameters.

\section{Conclusions}
\label{Secdiscuss}

In this paper we have shown that IIA string theory compactified on CY manifolds admits non-supersymmetric AdS or dS vacua where all the moduli are stabilised and the string coupling is exponentially small. The string scale and the supersymmetry breaking scale are schematically given by
\be
m_s \simeq \frac{g_s}{\sqrt{\Vol}} M_p \;, \;\; m_{\frac32} \simeq \frac{g_s^2|W_0|}{\Vol} M_p \;.
\ee 
Therefore the exponentially small coupling can naturally generate the TeV scale with $g_s \sim 10^{-7}$. This then leads to an intermediate string scale. It is worth noting that although the string coupling is exponentially small, the exponentially large complex-structure `volume' $\Vol'$ combines with $g_s$ to make the physical standard-model couplings, which are given by the four-dimensional superfields $S$ and $U_{\lambda}$,  of appropriate magnitude. 

Since these models are mirror to the LARGE-volume IIB compactifications, much of the phenomenological discussion based around those models will cross over to ours under the mirror map of section \ref{cycomwithout}. In particular the K\"ahler moduli inflation scenario of  \cite{Conlon:2005jm} simply maps to complex-structure moduli inflation (with $u_s$ the inflaton). In that sense this inflation scenario can be thought of as general type II inflation.

Although our compactifications can be identified as mirrors to existing IIB constructions, they can form a base for calculations and scenarios that are difficult to construct on the IIB side. We have already seen an example of this in section \ref{upsect} where the IIB dual to the uplifting term would be more difficult to construct. In general this hope applies more specifically to the matter sector constructions since for them the IIA/IIB duality is less understood. Perhaps being able to study the phenomenology of these models from both sides of the mirror will increase our understanding of them.   

 \vskip1.4cm

\subsection*{$\hskip5.8cm$Acknowledgements}

We thank Jose Juan Blanco-Pillado, Joe Conlon, Marta G\'omez-Reino, James Gray,  Maxim Pospelov,  Fernando Quevedo and Adam Ritz for useful discussions. 

EP is supported by a STFC Postdoctoral Fellowship. 
GT is supported by MEC and FEDER under grant FPA2006-05485, by  
CAM under grant HEPHACOS P-ESP-00346, and
by the ÒUniverseNetÓ network  (MRTN-CT-2006-035863).

\bigskip

\appendix

\section{The IIB Imaginary-Self-Dual conditions }\label{AppendixA}

In this appendix we show that the F-term equations in IIA, that we solved in the main part of the paper, are equivalent to the ISD condition in the IIB mirror theory.  
The IIB mirror set-up has the fluxes
\be
G_3 \equiv F_3 - iS H_3 = -f_0 \alpha_0 + \tilde{f}^i\alpha_i + f_i \beta^i - \tilde{f}_0 \beta^0 +i S h_0\,\beta^0 \;.
\ee
The signs of the fluxes are fixed so that they provide the correct expression for the superpotential, through the GVW formula $W=\int{G_3 \w \Omega}$. There is a subtlety here due to the fact that the IIB complex-structure moduli are defined with an opposite sign prepotential which corresponds to the minus sign in the interchange between the imaginary parts of the IIB complex-structure moduli and the K\"ahler moduli. In this appendix we work with the fields that are the direct mirrors so that their prepotential and gauge-kinetic matrix is given by (\ref{lvkpp}) and (\ref{gkpt}). We keep the same notation for the fields so that in this appendix $T^i$ are the IIB complex-structure fields.  
We want to impose the condition 
\be
G_3 = - i \star G_3 \;, \label{isdcondi}
\ee
and check that it provides the same conditions that come from the F-term equations on the IIA side. 
To analyse the above equation we consider integrals with respect to each component of  the basis which are given by \cite{Suzuki:1995rt,Andrianopoli:1996cm,Grimm:2004ua}
\ba
\int{\alpha_I \w \star \alpha_J} &=& -\left[ (\im{N}) + (\re{N})(\im{N^{-1}})(\re{N})  \right]_{IJ} \;, \nonumber \\
\int{\beta^I \w \star \beta^J} &=& - (\im N^{-1})^{IJ} \;, \nonumber \\
\int{\alpha_I \w \star \beta^J} &=& - \left[ (\re{N}) (\im{N}^{-1}) \right]^J_I \;.
\ea
The matrix $N$ is determined by the prepotential as in (\ref{gkpt}). We now go on to consider the solution to (\ref{isdcondi}) for the cases of the large complex-structure limit prepotential (\ref{lvkpp}) and the corrected one (\ref{corpot}), which should be equivalent to our analysis of the IIA F-terms without and with $\alpha'$ corrections respectively.

\subsection{The large complex-structure limit: the mirror to no $\alpha'$ corrections}
\label{appendixa1}

In this set-up the matrix $N$ reads \cite{Louis:2002ny}
\ba
\re{N} &=&  \left(  \begin{array}{cc} -\frac13 K_{ijk}b^ib^jb^k & \half K_{ijk} b^jb^k \\ \half K_{ijk}b^j b^k & - K_{ijk}b^k \end{array} \right)  \;, \\
\im{N} &=&  -\frac{\ka}{6} \left(  \begin{array}{cc}  1 +4 b^i b^j K^T_{ij}   &  -4 K^T_{ij} b^j\\ -4 K^T_{ij} b^j & 4 K^T_{ij} \end{array} \right)  \;, \\
\im{N}^{-1} &=&  -\frac{6}{\ka}\left(  \begin{array}{cc} 1 &b^i \\ b^i & b^ib^j + \frac14 (K^T)^{ij}\end{array} \right)  \;. \label{Nnoal}
\ea 
Now for the analysis we will use the definitions of the vectors $\tilde{e}_I $ and $\tilde{m}^I$, that we recall here 
\ba
\tilde{e}_I &=& \left( \tilde{f}_0 +  h_0 
\sigma, -f_i \right), \; \tilde{m}^I = \left( -f_0, \tilde{f}^i \right) \;. 
\ea
We want to check that the solution (\ref{bvev}), (\ref{TFT5}), (\ref{snoal}) solves the ISD equations, where we constrain the flux combination in (\ref{TFT5}) to vanish. For that solution we can use the following simple  identity, already mentioned in the main part of the paper
 \ba
 \label{simident}
 \left( \tilde{e}_I - \re{N}_{IJ} \tilde{m}^{J} \right) \,=\, 0\;\,\,.
 \ea
We proceed considering  the various integrals of condition (\ref{isdcondi}) with respect to each  basis component.
\begin{enumerate}
\item The integral with respect to $\beta^q$ leads to the condition
\ba
\tilde{f}^q
&=&i f_0\left[ (\re{N}) (\im{N}^{-1}) \right]^{q}_0
- i \tilde{f}^{j} \left[ (\re{N}) (\im{N}^{-1}) \right]^q_j\nonumber \\ &-&
i f_j\, (\im N^{-1})^{q j}+i \tilde{f}_0 
 (\im N^{-1})^{0 q} + S h_0  (\im N^{-1})^{0 q}\label{intbq}\,.
\ea
The real part of this equation reads
\be
\tilde{f}^q\,=\,-\frac{6\,s\,h_0}{\kappa}\,b^q \;,
\ee
that is satisfied for our configuration using  our solution for $s$. The imaginary part of (\ref{intbq}) can be written as
\be
\left(\tilde{e}_I- \re{N}_{I J}\tilde{m}^I\right)(\im N^{-1})^{I q}\,,
\ee
that vanishes using (\ref{simident}).
\item The integral  with respect to $\alpha_q$ leads to the condition
\ba
-f_q
&=&i f_0\left[ (\im{N}) + (\re{N})(\im{N^{-1}})(\re{N})
 \right]_{0q} \nonumber \\ &-&
 i \tilde{f}^{j} \left[
 (\im{N}) + (\re{N})(\im{N^{-1}})(\re{N})
 \right]_{qj}\nonumber \\ &-&
i f_j\, \left[
(\re{N}) (\im{N}^{-1})
\right]^{j}_q+i \tilde{f}_0 
 \left[
(\re{N}) (\im{N}^{-1})
\right]^{0}_q \nonumber \\ &+&
 S h_0  \left[
(\re{N}) (\im{N}^{-1})
\right]^{0}_q \label{intbq1} \;.
\ea
The real part of this condition is
\be
-f_q\,=\,s\, h_0\, \left[
(\re{N}) (\im{N}^{-1})
\right]^{0}_q \;.
\ee
Using (\ref{simident}), this can be written as
\be
\,(\re{N})_{qL}\,\tilde{m}^L\,=\,s\, h_0\, \left[(\re{N}) (\im{N}^{-1})
\right]^{0}_q \;,
\ee
or equivalently, using the fact that $(\re{N})_{ML}$ is invertible,
\be
\tilde{m}^L\,=\,s\, h_0\,  (\im{N}^{-1})^{L 0} \;. \label{reaq}
\ee
It is simple to check that this identity is indeed satisfied for our solution.  
The condition coming from the imaginary part can be conveniently reassembled in the following way
\ba 
0 &=& -\im{N}_{q J} \tilde{m}^J \nonumber \\
&+& 
 \left[(\re{N}) (\im{N}^{-1})
\right]^{L}_q\,\left\{
\tilde{e}_L- \re{N}_{L J}\tilde{m}^J
\right\} \;. \label{iaq}
\ea
The second line obviously vanishes using (\ref{simident}). The first
line gives
\be
0 \,=\, \frac{2\kappa\kappa_{iq}}{3}\left( f_0 b^i + \tilde{f}^i \right) \;,
\ee
which is satisfied for our solution.
\item The integral with respect to $\beta^0$ gives
the condition
\ba
-f_0
&=&i f_0\left[ (\re{N}) (\im{N}^{-1}) \right]^{0}_0
- i \tilde{f}^{j} \left[ (\re{N}) (\im{N}^{-1}) \right]^{0}_{j}\nonumber \\ &-&
i f_j\, (\im N^{-1})^{ j0}+i \tilde{f}_0 
 (\im N^{-1})^{00} + S h_0  (\im N^{-1})^{00}\label{intbq2} \;.
\ea
The real part reads
\be
f_0\,=\,-\,s\,h_0\,(\im N^{-1})^{00} \;,
\ee
that is satisfied for our solution. It  is also  immediate to show
that the imaginary part is equivalent to  (\ref{simident}).
\item
The integral with respect to $\alpha_0$ gives
the condition
\ba
\tilde{f}_0 - i S h_0&=&
i f_0\left[ (\im{N}) + (\re{N})(\im{N^{-1}})(\re{N})
 \right]_{00}\nonumber \\ &-&
 i \tilde{f}^{j} \left[
 (\im{N})  (\re{N})(\im{N^{-1}})(\re{N})
 \right]_{j0} - 
i f_j\, \left[
(\re{N}) (\im{N}^{-1})
\right]^{j}_0
\nonumber \\ &+&
i \tilde{f}_0 
 \left[
(\re{N}) (\im{N}^{-1})
\right]^{0}_0  
+ S h_0  \left[
(\re{N}) (\im{N}^{-1})
\right]^{0}_0 \label{intbq3} \;.
\ea
The real part of this is 
\be
\sigma h_0+\tilde{f}_0\,=\,-\frac{1}{\kappa}\,s\,h_0\,K_{ijk} b^i b^j b^k \;,
\ee
that is satisfied for our solution. The imaginary part is conveniently reassembled as
\ba 
-s h_0 &=&- \im{N}_{0 J} \tilde{m}^J \nonumber \\
&+& 
 \left[(\re{N}) (\im{N}^{-1})
\right]^{L}_0\,\left\{
\tilde{e}_L- \re{N}_{L J}\tilde{m}^J
\right\} \;.
\ea
The second line again vanishes using (\ref{simident}). The first line reads
\be
s h_0\,=\, \frac{\kappa}{6}\left[ 
\left(1 +4 b^i b^j \kappa_{ij}\right)\,f_0
+
4 \kappa_{ij} b^i \tilde{f}^j
\right] \;,
\ee
that is again satisfied for our solution. 
\end{enumerate}
This concludes the computation.

\subsection{Away from large complex-structure: the mirror to  $\alpha'$ corrections}
\label{AppendixA2}

In this case the gauge coupling matrix $N$, obtained from the corrected prepotential (\ref{corpot}), reads
\ba
\re{N} &=&  \left(  \begin{array}{cc} -\frac{1}{3} K_{ijk}b^ib^jb^k + \kappa_i b^i \Delta & \frac{1}{2} K_{ijk} b^jb^k+K_i^{(2)}-\frac{1}{2}\kappa_i \Delta  \label{Nwal1} \\ 
\frac{1}{2} K_{ijk}b^j b^k+K^{(2)}_i - \frac{1}{2}\kappa_i \Delta & - K_{ijk}b^k + 2 K^{(1)}_{ij} \end{array} \right)  \;, \\
\im{N} &=&  \left( \begin{array}{cc}  \tilde{g}_{ij}b^ib^j - \alpha   &  -\tilde{g}_{ij} b^j\\ - \tilde{g}_{ij} b^j & \tilde{g}_{ij} \end{array} \right)  \;, \\
\im{N}^{-1} &=&  \frac{1}{\alpha}\left(  \begin{array}{cc} -1 & -b^i \\ -b^i & -b^i b^j+\alpha \tilde{g}^{ij}\end{array} \right)  \;, \label{Nwal}
\ea
where we have defined the following
\begin{equation}
\Delta = 1-\frac{\kappa_{-6}}{\kappa_{-3/2}}, \hspace{1cm} \tilde{g}_{ij} = \kappa_{ij}-\frac{3 \kappa_i \kappa_j}{2 \kappa_{-3/2}}, \hspace{1cm} \alpha = \frac{\kappa_{-6} \kappa_3}{6 \kappa_{-3/2}}.
\end{equation}
For completeness we also include here the derivatives of the corrected K\"ahler potential
\be
K^T_{i\bar{j}} = -\frac{3}{2\kappa_3} \left( \kappa_{ij} -  \frac{3\kappa_i \kappa_j}{2\kappa_3} \right)  \hskip0.5cm, \hskip0.5cm (K^T)^{i\bar{j}} = -\frac{2\kappa_3}{3} \left( \kappa^{ij} - \frac{3\tau^i\tau^j}{\kappa_{-6}} \right)\;. \label{ecorkder}
\ee  

The solution we wish to recover is given by (\ref{bvev}), (\ref{kcorsol}), (\ref{scorsol}).
Using (\ref{Nwal1}-\ref{Nwal}) it can be checked that the identity (\ref{simident}) still holds. This means that much of the calculation will follow as in section \ref{appendixa1} and so we just present the results. 
\begin{enumerate}
\item
The first integral is with respect to $\beta^q$, and yields the following solution for the real component
\begin{equation}
\tilde f^q = -\frac{sh_0 b^q}{\alpha} \;,
\end{equation}
which is satisfied for our solution. The imaginary part vanishes using (\ref{simident}).
\item
Now consider the $\alpha_q$ integral. The real part reduces to (\ref{reaq}) which just gives 
\begin{equation}
h_0 s = \alpha f_0 \;.
\end{equation}
The imaginary part can be written again as in (\ref{iaq})  and gives
\begin{equation}
0 = \tilde{g}_{qi} \left( -b^if_0 - \tilde{f}^i \right) \;, 
\end{equation}
which is satisfied.
\item
The third integral is with respect to $\beta^0$ and we see that the real condition reduces to the usual $h_0 s = \alpha f_0$. The imaginary part again vanishes due to (\ref{simident}).
\item
The last integral is with respect to $\alpha_0$, and this gives us the real piece
\begin{equation}
\tilde f_0 + \sigma h_0 = -\frac{s h_0}{\alpha} \left(\frac{K_{ijk}b^i b^j b^k}{6} + K^{(2)}_i b^i + \frac{\kappa_i b^i \Delta}{2} \right) \;,
\end{equation}
which can be checked to be satisfied for our solution. The imaginary part again becomes the familiar $s h_0 = f_0 \alpha$. 
\end{enumerate}
This concludes the check on the ISD conditions away from the large complex-structure limit.




\begin{thebibliography}{10}

{\small 

\bibitem{Giddings:2001yu}
  S.~B.~Giddings, S.~Kachru and J.~Polchinski,
  ``Hierarchies from fluxes in string compactifications,''
  Phys.\ Rev.\  D {\bf 66} (2002) 106006
  [arXiv:hep-th/0105097].

\bibitem{Kachru:2003aw}
  S.~Kachru, R.~Kallosh, A.~Linde and S.~P.~Trivedi,
  ``De Sitter vacua in string theory,''
  Phys.\ Rev.\  D {\bf 68} (2003) 046005
  [arXiv:hep-th/0301240].

\bibitem{Koerber:2007xk}
  P.~Koerber and L.~Martucci,
  ``From ten to four and back again: how to generalize the geometry,''
  JHEP {\bf 0708} (2007) 059
  [arXiv:0707.1038 [hep-th]].

\bibitem{Behrndt:2004km}
  K.~Behrndt and M.~Cvetic,
  ``General N = 1 supersymmetric flux vacua of (massive) type IIA string
  theory,''
  Phys.\ Rev.\ Lett.\  {\bf 95} (2005) 021601
  [arXiv:hep-th/0403049].
  
\bibitem{Behrndt:2004mj}
  K.~Behrndt and M.~Cvetic,
  ``General N = 1 supersymmetric fluxes in massive type IIA string theory,''
  Nucl.\ Phys.\  B {\bf 708} (2005) 45
  [arXiv:hep-th/0407263].

\bibitem{Lust:2004ig}
  D.~Lust and D.~Tsimpis,
  ``Supersymmetric AdS(4) compactifications of IIA supergravity,''
  JHEP {\bf 0502} (2005) 027
  [arXiv:hep-th/0412250].

\bibitem{Kachru:2002sk}
  S.~Kachru, M.~B.~Schulz, P.~K.~Tripathy and S.~P.~Trivedi,
  ``New supersymmetric string compactifications,''
  JHEP {\bf 0303} (2003) 061
  [arXiv:hep-th/0211182].

\bibitem{Villadoro:2005cu}
  G.~Villadoro and F.~Zwirner,
  ``N = 1 effective potential from dual type-IIA D6/O6 orientifolds with
  general fluxes,''
  JHEP {\bf 0506} (2005) 047
  [arXiv:hep-th/0503169].

\bibitem{House:2005yc}
  T.~House and E.~Palti,
  ``Effective action of (massive) IIA on manifolds with SU(3) structure,''
  Phys.\ Rev.\  D {\bf 72}, 026004 (2005)
  [arXiv:hep-th/0505177].

\bibitem{Micu:2006ey}
  A.~Micu, E.~Palti and P.~M.~Saffin,
  ``M-theory on seven-dimensional manifolds with SU(3) structure,''
  JHEP {\bf 0605}, 048 (2006)
  [arXiv:hep-th/0602163].

\bibitem{Ihl:2007ah}
  M.~Ihl, D.~Robbins and T.~Wrase,
  ``Toroidal Orientifolds in IIA with General NS-NS Fluxes,''
  JHEP {\bf 0708} (2007) 043
  [arXiv:0705.3410 [hep-th]].
  
\bibitem{Grana:2006kf}
  M.~Grana, R.~Minasian, M.~Petrini and A.~Tomasiello,
  ``A scan for new N=1 vacua on twisted tori,''
  JHEP {\bf 0705}, 031 (2007)
  [arXiv:hep-th/0609124].

\bibitem{Aldazabal:2007sn}
  G.~Aldazabal and A.~Font,
  ``A second look at N=1 supersymmetric AdS4 vacua of type IIA supergravity,''
  JHEP {\bf 0802} (2008) 086
  [arXiv:0712.1021 [hep-th]].

\bibitem{Tomasiello:2007eq}
  A.~Tomasiello,
  ``New string vacua from twistor spaces,''
  arXiv:0712.1396 [hep-th].

\bibitem{Koerber:2008rx}
  P.~Koerber, D.~Lust and D.~Tsimpis,
  ``Type IIA AdS4 compactifications on cosets, interpolations and domain
  walls,''
  arXiv:0804.0614 [hep-th].

\bibitem{fluxrev}
  M.~Grana,
  ``Flux compactifications in string theory: A comprehensive review,''
  Phys.\ Rept.\  {\bf 423} (2006) 91
  [arXiv:hep-th/0509003];

  M.~R.~Douglas and S.~Kachru,
  ``Flux compactification,''
  Rev.\ Mod.\ Phys.\  {\bf 79} (2007) 733
  [arXiv:hep-th/0610102].

\bibitem{Gurrieri:2002wz}
  S.~Gurrieri, J.~Louis, A.~Micu and D.~Waldram,
  ``Mirror symmetry in generalized Calabi-Yau compactifications,''
  Nucl.\ Phys.\  B {\bf 654} (2003) 61
  [arXiv:hep-th/0211102].

\bibitem{Shelton:2005cf}
  J.~Shelton, W.~Taylor and B.~Wecht,
  ``Nongeometric flux compactifications,''
  JHEP {\bf 0510} (2005) 085
  [arXiv:hep-th/0508133].

\bibitem{Camara:2005dc}
  P.~G.~Camara, A.~Font and L.~E.~Ibanez,
  ``Fluxes, moduli fixing and MSSM-like vacua in a simple IIA orientifold,''
  JHEP {\bf 0509} (2005) 013
  [arXiv:hep-th/0506066].

\bibitem{Aldazabal:2006up}
  G.~Aldazabal, P.~G.~Camara, A.~Font and L.~E.~Ibanez,
  ``More dual fluxes and moduli fixing,''
  JHEP {\bf 0605} (2006) 070
  [arXiv:hep-th/0602089].

\bibitem{Benmachiche:2006df}
  I.~Benmachiche and T.~W.~Grimm,
  ``Generalized N = 1 orientifold compactifications and the Hitchin
  functionals,''
  Nucl.\ Phys.\  B {\bf 748} (2006) 200
  [arXiv:hep-th/0602241].

\bibitem{Grana:2006hr}
  M.~Grana, J.~Louis and D.~Waldram,
  ``SU(3) x SU(3) compactification and mirror duals of magnetic fluxes,''
  JHEP {\bf 0704} (2007) 101
  [arXiv:hep-th/0612237].

\bibitem{Micu:2007rd}
  A.~Micu, E.~Palti and G.~Tasinato,
  ``Towards Minkowski vacua in type II string compactifications,''
  JHEP {\bf 0703} (2007) 104
  [arXiv:hep-th/0701173].

\bibitem{Palti:2007pm}
  E.~Palti,
  ``Low Energy Supersymmetry from Non-Geometry,''
  JHEP {\bf 0710} (2007) 011
  [arXiv:0707.1595 [hep-th]].

\bibitem{Wecht:2007wu}
  B.~Wecht,
  ``Lectures on Nongeometric Flux Compactifications,''
  Class.\ Quant.\ Grav.\  {\bf 24}, S773 (2007)
  [arXiv:0708.3984 [hep-th]].

\bibitem{Balasubramanian:2005zx}
  V.~Balasubramanian, P.~Berglund, J.~P.~Conlon and F.~Quevedo,
  ``Systematics of moduli stabilisation in Calabi-Yau flux
  compactifications,''
  JHEP {\bf 0503} (2005) 007
  [arXiv:hep-th/0502058].
  
\bibitem{Conlon:2005ki}
  J.~P.~Conlon, F.~Quevedo and K.~Suruliz,
  ``Large-volume flux compactifications: Moduli spectrum and D3/D7 soft
  supersymmetry breaking,''
  JHEP {\bf 0508} (2005) 007
  [arXiv:hep-th/0505076].

  J.~P.~Conlon,
  ``The QCD axion and moduli stabilisation,''
  JHEP {\bf 0605} (2006) 078
  [arXiv:hep-th/0602233].

  J.~P.~Conlon, S.~S.~Abdussalam, F.~Quevedo and K.~Suruliz,
  ``Soft SUSY breaking terms for chiral matter in IIB string
  compactifications,''
  JHEP {\bf 0701} (2007) 032
  [arXiv:hep-th/0610129].

  J.~P.~Conlon and D.~Cremades,
  ``The neutrino suppression scale from large volumes,''
  Phys.\ Rev.\ Lett.\  {\bf 99} (2007) 041803
  [arXiv:hep-ph/0611144].

  J.~P.~Conlon,
  ``Mirror Mediation,''
  arXiv:0710.0873 [hep-th].

\bibitem{Conlon:2005jm}
  J.~P.~Conlon and F.~Quevedo,
  ``Kaehler moduli inflation,''
  JHEP {\bf 0601} (2006) 146
  [arXiv:hep-th/0509012].

\bibitem{McAllister:2007bg}
  S.~H.~Henry Tye,
  ``Brane inflation: String theory viewed from the cosmos,''
  arXiv:hep-th/0610221; 

  R.~Kallosh,
  ``On Inflation in String Theory,''
  Lect.\ Notes Phys.\  {\bf 738} (2008) 119
  [arXiv:hep-th/0702059]; 

  C.~P.~Burgess,
  ``Lectures on Cosmic Inflation and its Potential Stringy Realizations,''
  PoS {\bf P2GC}, 008 (2006)
  [Class.\ Quant.\ Grav.\  {\bf 24}, S795 (2007)]
  [arXiv:0708.2865 [hep-th]];

  L.~McAllister and E.~Silverstein,
  ``String Cosmology: A Review,''
  Gen.\ Rel.\ Grav.\  {\bf 40} (2008) 565
  [arXiv:0710.2951 [hep-th]].

\bibitem{Hertzberg:2007ke}
  M.~P.~Hertzberg, M.~Tegmark, S.~Kachru, J.~Shelton and O.~Ozcan,
  ``Searching for Inflation in Simple String Theory Models: An Astrophysical
  Perspective,''
  Phys.\ Rev.\  D {\bf 76} (2007) 103521
  [arXiv:0709.0002 [astro-ph]].

\bibitem{Hertzberg:2007wc}
  M.~P.~Hertzberg, S.~Kachru, W.~Taylor and M.~Tegmark,
  ``Inflationary Constraints on Type IIA String Theory,''
  JHEP {\bf 0712} (2007) 095
  [arXiv:0711.2512 [hep-th]].

\bibitem{Acharya:2006ne}
  B.~S.~Acharya, F.~Benini and R.~Valandro,
  ``Fixing moduli in exact type IIA flux vacua,''
  JHEP {\bf 0702} (2007) 018
  [arXiv:hep-th/0607223].

\bibitem{Candelas:1990pi}
  P.~Candelas and X.~de la Ossa,
  ``MODULI SPACE OF CALABI-YAU MANIFOLDS,''
  Nucl.\ Phys.\  B {\bf 355} (1991) 455.


\bibitem{Grimm:2004ua}
  T.~W.~Grimm and J.~Louis,
  ``The effective action of type IIA Calabi-Yau orientifolds,''
  Nucl.\ Phys.\  B {\bf 718} (2005) 153
  [arXiv:hep-th/0412277].

\bibitem{Andrianopoli:1996cm}
  L.~Andrianopoli, M.~Bertolini, A.~Ceresole, R.~D'Auria, S.~Ferrara, P.~Fre and T.~Magri,
  ``N = 2 supergravity and N = 2 super Yang-Mills theory on general scalar
  manifolds: Symplectic covariance, gaugings and the momentum map,''
  J.\ Geom.\ Phys.\  {\bf 23} (1997) 111
  [arXiv:hep-th/9605032].

\bibitem{Candelas:1990rm}
  P.~Candelas, X.~C.~De La Ossa, P.~S.~Green and L.~Parkes,
  ``A pair of Calabi-Yau manifolds as an exactly soluble superconformal
  theory,''
  Nucl.\ Phys.\  B {\bf 359} (1991) 21.
 
\bibitem{DeWolfe:2005uu}
  O.~DeWolfe, A.~Giryavets, S.~Kachru and W.~Taylor,
  ``Type IIA moduli stabilization,''
  JHEP {\bf 0507} (2005) 066
  [arXiv:hep-th/0505160].

\bibitem{Denef:2004dm}
  F.~Denef, M.~R.~Douglas and B.~Florea,
  ``Building a better racetrack,''
  JHEP {\bf 0406} (2004) 034
  [arXiv:hep-th/0404257].

\bibitem{Grimm:2005fa}
  T.~W.~Grimm,
  ``The effective action of type II Calabi-Yau orientifolds,''
  Fortsch.\ Phys.\  {\bf 53} (2005) 1179
  [arXiv:hep-th/0507153].
  
  T.~W.~Grimm and J.~Louis,
  ``The effective action of type IIA Calabi-Yau orientifolds,''
  Nucl.\ Phys.\  B {\bf 718} (2005) 153
  [arXiv:hep-th/0412277].

\bibitem{Becker:2002nn}
  K.~Becker, M.~Becker, M.~Haack and J.~Louis,
  ``Supersymmetry breaking and alpha'-corrections to flux induced
  potentials,''
  JHEP {\bf 0206} (2002) 060
  [arXiv:hep-th/0204254].

\bibitem{Blumenhagen:2007sm}
  R.~Blumenhagen, S.~Moster and E.~Plauschinn,
  ``Moduli Stabilisation versus Chirality for MSSM like Type IIB
  Orientifolds,''
  JHEP {\bf 0801} (2008) 058
  [arXiv:0711.3389 [hep-th]].

\bibitem{Dine:1986vd}
  M.~Dine and N.~Seiberg,
  ``Nonrenormalization Theorems in Superstring Theory,''
  Phys.\ Rev.\ Lett.\  {\bf 57} (1986) 2625.

\bibitem{Berg:2005ja}
  M.~Berg, M.~Haack and B.~Kors,
  ``String loop corrections to Kaehler potentials in orientifolds,''
  JHEP {\bf 0511} (2005) 030
  [arXiv:hep-th/0508043].

\bibitem{Berg:2007wt}
  M.~Berg, M.~Haack and E.~Pajer,
  ``Jumping Through Loops: On Soft Terms from Large Volume Compactifications,''
  JHEP {\bf 0709} (2007) 031
  [arXiv:0704.0737 [hep-th]].

\bibitem{Cicoli:2007xp}
  M.~Cicoli, J.~P.~Conlon and F.~Quevedo,
  ``Systematics of String Loop Corrections in Type IIB Calabi-Yau Flux
  Compactifications,''
  JHEP {\bf 0801} (2008) 052
  [arXiv:0708.1873 [hep-th]].

\bibitem{Angelantonj:2002ct}
  C.~Angelantonj and A.~Sagnotti,
  ``Open strings,''
  Phys.\ Rept.\  {\bf 371} (2002) 1
  [Erratum-ibid.\  {\bf 376} (2003) 339]
  [arXiv:hep-th/0204089].

\bibitem{Looyestijn:2008pg}
  H.~Looyestijn and S.~Vandoren,
  ``On NS5-brane instantons and volume stabilization,''
  arXiv:0801.3949 [hep-th].

\bibitem{Silverstein:2007ac}
  E.~Silverstein,
  ``Simple de Sitter Solutions,''
  arXiv:0712.1196 [hep-th].

\bibitem{Villadoro:2006ia}
  G.~Villadoro and F.~Zwirner,
  ``D terms from D-branes, gauge invariance and moduli stabilization in  flux
  compactifications,''
  JHEP {\bf 0603} (2006) 087
  [arXiv:hep-th/0602120].

\bibitem{Burgess:2003ic}
  C.~P.~Burgess, R.~Kallosh and F.~Quevedo,
  ``de Sitter string vacua from supersymmetric D-terms,''
  JHEP {\bf 0310} (2003) 056
  [arXiv:hep-th/0309187].

\bibitem{Achucarro:2006zf}
  A.~Achucarro, B.~de Carlos, J.~A.~Casas and L.~Doplicher,
  ``de Sitter vacua from uplifting D-terms in effective supergravities from
  realistic strings,''
  JHEP {\bf 0606} (2006) 014
  [arXiv:hep-th/0601190].

\bibitem{Cremades:2007ig}
  D.~Cremades, M.~P.~Garcia del Moral, F.~Quevedo and K.~Suruliz,
  ``Moduli stabilisation and de Sitter string vacua from magnetised D7
  branes,''
  JHEP {\bf 0705} (2007) 100
  [arXiv:hep-th/0701154].

\bibitem{Becker:1995kb}
  K.~Becker, M.~Becker and A.~Strominger,
  ``Five-Branes, Membranes And Nonperturbative String Theory,''
  Nucl.\ Phys.\  B {\bf 456} (1995) 130
  [arXiv:hep-th/9507158].

\bibitem{Freed:1999vc}
  D.~S.~Freed and E.~Witten,
  ``Anomalies in string theory with D-branes,''
  arXiv:hep-th/9907189.

\bibitem{LoaizaBrito:2006se}
  O.~Loaiza-Brito,
  ``Freed-Witten anomaly in general flux compactification,''
  Phys.\ Rev.\  D {\bf 76} (2007) 106015
  [arXiv:hep-th/0612088].
 
\bibitem{Suzuki:1995rt}
  H.~Suzuki,
  ``Calabi-Yau compactification of type IIB string and a mass formula of the
  extreme black holes,''
  Mod.\ Phys.\ Lett.\  A {\bf 11} (1996) 623
  [arXiv:hep-th/9508001].

\bibitem{Louis:2002ny}
  J.~Louis and A.~Micu,
  ``Type II theories compactified on Calabi-Yau threefolds in the presence  of
  background fluxes,''
  Nucl.\ Phys.\  B {\bf 635} (2002) 395
  [arXiv:hep-th/0202168].
   
 }

\end{thebibliography}
\end{document}